\documentclass[aps,prl,reprint,amsmath,amssymb,superscriptaddress]{revtex4-1}

\usepackage{graphicx}
\usepackage{dcolumn}
\usepackage{bm}
\usepackage{hyperref}
\usepackage{aas_macros}
\usepackage{xcolor}
\usepackage{braket}

\usepackage{comment}

\usepackage{verbatim}

\newcommand{\orcid}[1]{\begingroup
  \hypersetup{hidelinks}\href{https://orcid.org/#1}{\includegraphics[width=10pt]{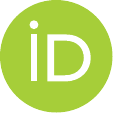}} \endgroup}

\begin{document}


\title{
Constraining Light QCD Axions with Isolated Neutron Star Cooling
}

\author{Antonio Gómez-Bañón \orcid{0009-0007-2546-5372}}
\affiliation{Departament de Física Aplicada, Universitat d’Alacant, 03690 Alicante, Spain}
\author{Kai Bartnick\orcid{0009-0001-8726-8485}}
\affiliation{Physik-Department, Technische Universit\"at M\"unchen,  85748 Garching, Germany}
\author{Konstantin Springmann\orcid{0000-0002-9617-6136}}
\affiliation{Physik-Department, Technische Universit\"at M\"unchen,  85748 Garching, Germany}
\affiliation{Department of Particle Physics and Astrophysics,
Weizmann Institute of Science, Rehovot 7610001, Israel}
\author{Jos\'e A. Pons \orcid{0000-0003-1018-8126}}
\affiliation{Departament de Física Aplicada, Universitat d’Alacant, 03690 Alicante, Spain}
\date{\today}

\begin{abstract}
The existence of light QCD axions, whose mass depends on an additional free parameter, can lead to a new ground state of matter, where the sourced axion field reduces the nucleon effective mass. 
The presence of the axion field has structural consequences, in particular, it results in a thinner (or even prevents its existence) heat-blanketing envelope, significantly altering the cooling patterns of neutron stars. We exploit the anomalous cooling behavior to constrain previously uncharted regions of the axion parameter space by comparing model predictions with existing data from isolated neutron stars. Notably, this
analysis does not require the light QCD axion to be the dark matter candidate.
\end{abstract}
\maketitle

\textit{Introduction}---There has been renewed interest in the physics of the QCD axion \cite{Peccei_1977a, Peccei_1977b, Wilczek_1978, Weinberg_1978}, both experimentally and theoretically. 
Despite decades of experimental effort, astrophysical probes remain the most sensitive \cite{Raffelt_2008,Caputo_2024}, with neutron star (NS) and supernova cooling providing some of the strongest bounds \cite{Page_2010,Fischer_2016,Leinson_2014,Sedrakian_2016,Sedrakian2019,Chang_2018,Hamaguchi_2018,Carenza_2019,Leinson_2021, Buschmann_2021}.
In this Letter, we investigate isolated NS cooling in the presence of QCD axions with mass lighter than predicted by QCD at a given $f_a$.
In dense astrophysical objects, their mass becomes tachyonic as first discovered in \cite{Hook:2017psm}, see also \cite{Balkin:2020dsr}.
The axion can then be sourced within the dense object, which in turn leads to a backreaction, as first discussed in \cite{Balkin:2022qer,Balkin:2023xtr}.

The defining coupling of QCD axions, the pseudo-Nambu-Goldstone boson $a(x)$ of a global axial $U(1)_{\text{PQ}}$, is 
their interaction with gluons,
\begin{equation}\label{eq:2gluoncoup}
    \mathcal{L}=\frac{g_s^2}{32\pi^2}\frac{a}{f_a}G_{\mu\nu}\Tilde{G}^{\mu\nu},
\end{equation}
where $g_s$ is the strong coupling, $G_{\mu\nu}$ denotes the gluon field strength, and $f_a$ is the axion decay constant. 
At energies below the QCD scale, nonperturbative effects break the shift symmetry and generate a periodic potential, which  takes the form~\cite{Cortona_2016}
\begin{equation}
    V(a)=-\epsilon m_\pi^2 f_\pi^2\left[\sqrt{1-\beta\sin^2\left(\frac{a}{2f_a}\right)}-1\right],
\end{equation}
where $m_\pi$ and $f_\pi$ are the pion mass and decay constant respectively, $\beta=4z/(1+z)^2$ with $z=m_u/m_d$ and $\epsilon$ is a model dependent parameter.
Here, $\epsilon=1$ corresponds to the potential of the standard QCD-axion.
We focus on the phenomenologically interesting case where $\epsilon\leq 1$, reducing the axion mass, which technically natural UV completions can achieve \cite{Hook:2017psm,Hook_2018b,DiLuzio:2021pxd}. 

The breaking of the shift symmetry also introduces a nonderivative coupling to nucleons, which reads \cite{Balkin:2022qer,Balkin:2023xtr,Springmann:2024mjp}
\begin{equation}
\label{eq:axion_nucleon_int}
\mathcal{L}\simeq-\sigma_N\bar{N}N\left[\sqrt{1-\beta\sin^2\left(\frac{a}{2f_a}\right)}-1\right],
\end{equation}
where $N=(p,n)^{\operatorname{T}}$ is the nucleon field, $\sigma_N\simeq 50~\text{MeV}$ is the nucleon sigma term.

The nonderivative coupling in Eq.~(\ref{eq:axion_nucleon_int}), while typically negligible for energy-loss arguments \cite{Buschmann_2021, raffelt1996stars}, can source the axion. In the presence of a nonzero baryon scalar density $n_s\equiv\braket{\bar{N}N}\neq0$, this term contributes to the potential and, above critical densities $n_s > n_s^c \equiv \epsilon m_\pi^2f_\pi^2 / \sigma_N $, new minima appear at $a=\pm\pi f_a$ \cite{Hook:2017psm,Balkin:2020dsr,Balkin:2021zfd,Balkin:2021wea,Balkin:2022qer,Balkin:2023xtr,Anzuini_2024,Gao_2022,Ramadan:2024vfc}.

The fact that astrophysical objects, such as the Earth, Sun, white dwarfs, or NSs, can source the axion has been used to put stringent and robust constraints on the axion parameter space \cite{Hook:2017psm,Huang_2019,Balkin:2022qer}.
Here, we follow this line of thought and investigate NS cooling in the presence of an axion condensate.

NS cooling offers insights into dense matter and particle physics. Early works \cite{Tsuruta_1965, Bahcall_1965} laid the groundwork following initial x-ray emission detections from NS surfaces \cite{Gursky_1963}. Recent simulations \cite{Pons_2019, DeGrandis_2021, Ascenzi_2024} continue efforts to link thermal emissions from isolated NSs to their interior properties \cite{Yakovlev_2004, Page_2004, Potekhin_2015}. 
Despite the limited availability of sources and some uncertainties in data, advancements in x-ray satellites over the past two decades have significantly improved observational capabilities and increased the number of available sources \cite{Vigano_2013, Kurpas_2023}. The standard theory explains most sources \cite{Page_2004}, but outliers like magnetars \cite{Vigano_2013} and exceptionally cool sources \cite{Marino_2024} challenge current understanding and offer opportunities to explore new physics.

Central to NS cooling studies is an energy balance equation combined with an envelope model \( T_s(T_b) \), relating the isothermal core temperature \( T_b \) to the surface temperature \( T_s \). 
For typical NS equations of state (EOS), envelope densities drop from \( (10^8-10^{10})~\text{g}/\text{cm}^3 \) to 0 over scales of approximately 100 m \cite{Beznogov_2021}.

The presence of a sourced QCD axion can lead to the formation of a new ground state (NGS) of nuclear matter, reminiscent of strange quark matter \cite{Witten_1984}.
Because of the mass reduction, there is a nonzero binding energy at finite densities $n_s^\ast>n_s^c$. 
This affects the stellar composition of compact stars \cite{Balkin:2022qer,Balkin:2023xtr}:
the density profile at the edge of the star drops from $n_s^\ast$  to 0 on scales dictated by the inverse in-medium axion mass $m_a(n_s)$, which can be much smaller than for typical envelopes \cite{Balkin:2023xtr}.
In the low $f_a$ limit, the energy density can drop from values much above $10^{10}~\text{g}/\text{cm}^3$ to 0 on infinitesimal scales. 

Consequently, the surface temperature is effectively equivalent to the core temperature  $T_b=T_s$, which leads to many orders of magnitude faster cooling. 
Similar effects have been used to argue against the existence of bare strange stars \cite{Blaschke2001}. 
For larger axion mass scales, the size of the envelope is reduced as a function of $f_a$ and the NS can be significantly less isolated, leading to faster cooling.

\textit{Neutron star envelopes}---Since the role of the NS envelope is central to this study, let us briefly review the procedure for constructing an envelope model. 
Its structure is determined by solving the hydrostatic equilibrium equation together with the energy flux equation. Denoting radial derivatives with primes, the hydrostatic equilibrium equation is 
\begin{equation}
 p' = - \frac{G M \rho}{r^2} \equiv - g \rho, \label{eq:HydrostaticEquilibrium}
\end{equation}
where $p$ and $\rho$ are the pressure and mass density of the fluid, $G$ is Newton's constant, $M$ is the mass of the star and we denote by $g$ the local gravity.
The energy flux $F_e$ in terms of the temperature gradient $T'$ is given by
\begin{equation}
 F_e = - {\cal K} T' , \label{eq:EnergyFlux}
\end{equation}
where ${\cal K}(\rho, T)$ is the thermal conductivity. 
Note that, in these two equations, we omitted general relativity corrections for clarity, which are, however, included in all calculations throughout this Letter. 
The two main assumptions for the thin layer of the envelope are that gravity is almost constant (since very little mass is contained in the outer layers) and the energy flux is nearly constant throughout the envelope (as there are no energy sources or sinks) and given by the Stefan-Boltzmann law at a given 
surface temperature ($T_s$),
$F_e = \sigma T_s^4$. 
With these assumptions, one can combine Eqs.~(\ref{eq:HydrostaticEquilibrium}) and (\ref{eq:EnergyFlux}) into
\begin{equation}
\frac{\partial T}{\partial p} = \frac{\sigma T_s^4}{{\cal K} \rho g }~.
\label{eq:dTdP}
\end{equation}
Hence, the pressure drop is connected to the drop in temperature, assuming hydrostatic equilibrium and constant flux. The difficulty lies in the complex nonlinear dependence of the thermal conductivity on $\rho$ and $T$. 
By integrating from the base of the envelope, typically defined at $\rho=\rho_b \equiv 10^{10}~\mathrm{g/cm^3}$ to the surface, Gudmundsson \cite{Gudmundsson_1982,Gudmundsson_1983} obtained a family of models and provided a simple fit valid in most typical NS conditions,
\begin{eqnarray}\label{eq:gudrel}
{T_{s6}}^4 = {g_{14}} (0.78~ {T_{b8}})^{2.2},
\end{eqnarray}
where $T_{b8}$ is the temperature at the base of the envelope $T_b$ in units of $10^8~\text{K}$, $T_{s6}$ is $T_s$ in units of $10^6~\text{K}$, and $g_{14}$ is the gravity in units of
$10^{14}~\mathrm{cm/s^2}$. 
As noted in the seminal paper by Gudmundsson \cite{Gudmundsson_1982}, for a
given opacity and EOS, $T_b$ is simply a function of $T_s^4/g$ since the dependence on the surface boundary condition is very weak. We will use this property below.

\textit{Neutron star envelopes with axion}---The presence of a sourced axion can drastically alter this picture.
An axion condensate can lead to an NGS of nuclear matter, as has been worked out in \cite{Balkin:2022qer,Balkin:2023xtr}, using a free Fermi gas EOS.
The free Fermi gas approximation, while certainly simplistic, captures the main features of this mechanism.
For this work, we employ the BSk26 EOS \cite{BSk26_2013} and include the effect of a sourced axion as a reduction of the energy per particle [by $\sim32\,\text{MeV}$; see Eq.~(\ref{eq:axion_nucleon_int})] and by adding (subtracting) its potential from the mass density (pressure).
This neglects relativistic corrections and modifications of the nuclear interaction, which is, however, a very good approximation at low densities.

We find that a NGS of nuclear matter appears for $\epsilon \lesssim 0.1$, which corresponds to critical densities \hbox{$n_s^c \lesssim 0.04~\mathrm{fm^{-3}}$}, 1 order below nuclear saturation density ($0.153~\mathrm{fm^{-3}}$) (see Appendix for details).
We explicitly verified that this is a general effect, occurring for various EOS, in particular within the framework of relativistic mean-field theory. Although the use of different EOS may slightly alter our results by $\mathcal{O}(1)$, we expect that the fundamental aspects remain unchanged.
\begin{figure}[ht]
    \centering
    \includegraphics[width=\linewidth]{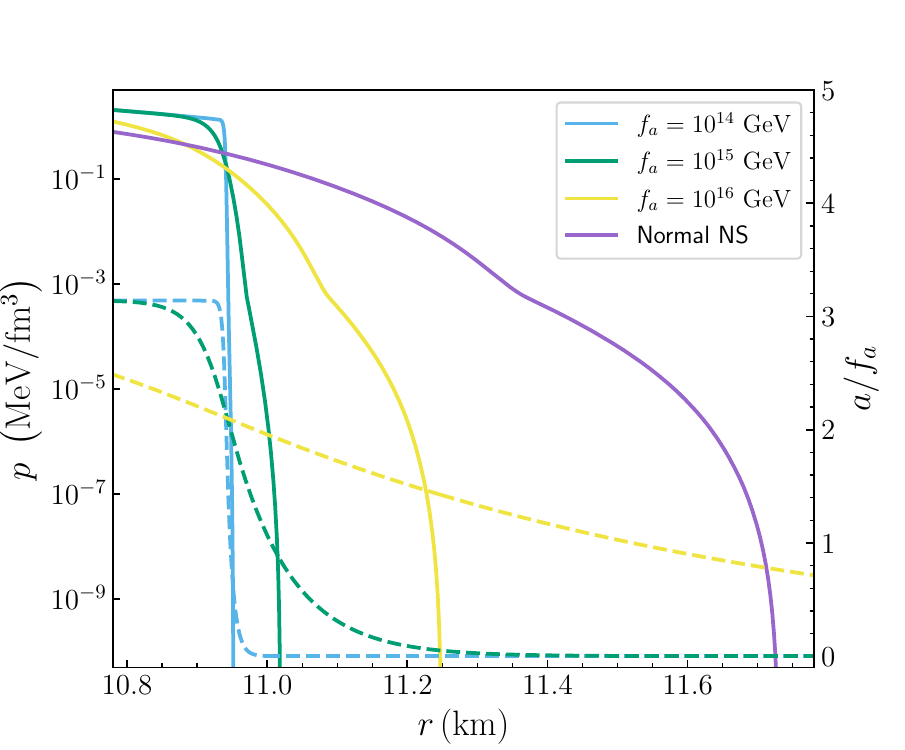}
    \caption{Matter pressure profiles (solid lines) and axion field profiles ($a/f_a$ in dashed lines) for stars in hydrostatic equilibrium with the same central pressure. We show the profiles in the outermost 1 km for $\epsilon=0.1$.}
    \label{fig:Pdrop}
\end{figure}

Including relativistic corrections and the axion gradient effects, the hydrostatic equilibrium equations known as Tolman-Oppenheimer-Volkoff equations \cite{Tolman_1939,Oppenheimer_1939} and the axion equation of motion form a coupled system of equations that has been derived in \cite{Balkin:2022qer,Balkin:2023xtr}. For the employed EOS, we start from a baseline model (without axion) resulting in a NS with central pressure of $100 ~\mathrm{MeV/fm^3}$ with $R_{\mathrm{NS}}=11.7~\mathrm{km}$ and $M_{\mathrm{NS}}=1.53~\mathrm{M_{\odot}}$.
To understand the structure of an NS in the NGS and an axion gradient-dominated envelope, it is instructive to write the axion contributions to the hydrostatic equilibrium as follows:
\begin{equation}
p' = - \rho (g + g_a),
\end{equation}
where $p$ is the pressure of the fluid, $g_a$ is given by
\begin{equation}\label{eq:gphi}
 g_a \equiv \frac{\partial \log m_N^{*}(a)}{\partial a}  a' =-\frac{1}{\rho}\frac{n_s}{n_s^c}\frac{\partial V}{\partial a}a'.
\end{equation}
Here, the axion gradient acts as an additional effective gravity (or equivalently, as surface tension) around the density where the axion field changes from $\pi f_a$ to $0$. 

\begin{figure}
    \centering
    \includegraphics[width=\linewidth]{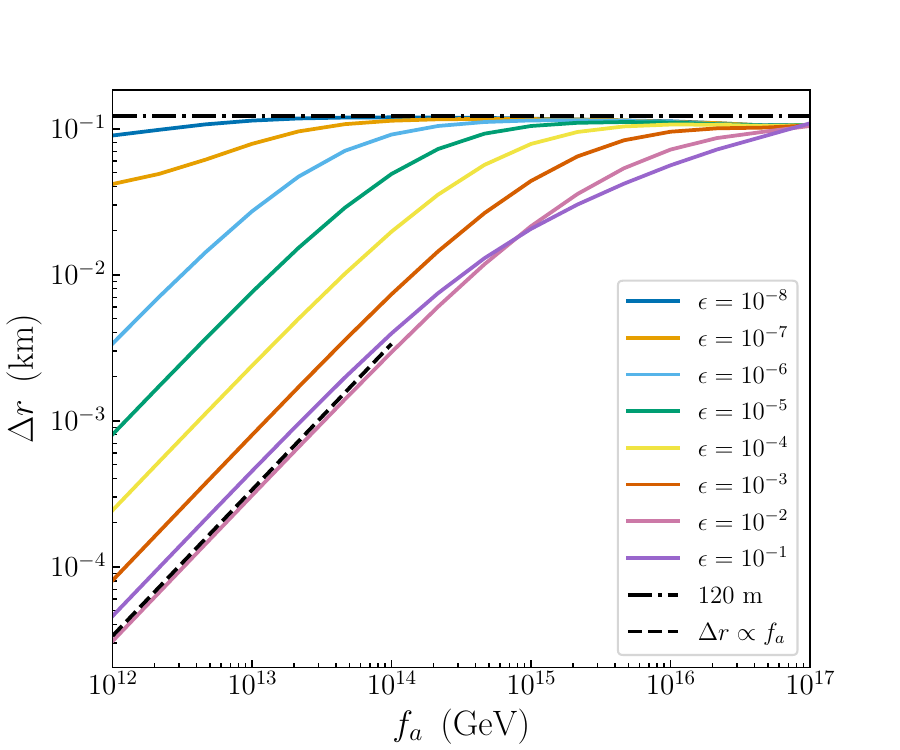}
    \caption{Scaling of the envelope size for different $\epsilon$ as a function of $f_a$. We define the envelope as the region where $\rho<\rho_b$. In the NGS case, and for sufficiently low $f_a$, the size of the envelope scales linearly with $f_a$ (black dashed line). The envelope size of the baseline model is represented by the black dash-dotted line.}
    \label{fig:envelope_scaling}
\end{figure}

For our baseline model, we solve the regular Tolman-Oppenheimer-Volkoff equations and find the thickness of the envelope to be $\Delta r=|r(\rho_b)-r_s|\simeq120~\mathrm{m}$.
Including the axion, we solve the coupled system numerically by employing a shooting method.
In the NGS case, the axion transition from $\pi f_a$ to 0 happens at the surface of the star and the outermost layer gets compressed.
This effect is shown in Fig.~\ref{fig:Pdrop}, where we compare NS envelope profiles for several values of $f_a$.
As $f_a$ is lowered, the larger $g_{a}$ results in the compression of the low-density part of the star and a gradually increasing pressure drop at $n_s \sim n_s^c$. A similar behavior is found in the energy density profiles.
We find that at low $f_a$ the envelope size scales as $\Delta r\propto f_a$, while for large $f_a$ we approach the constant $\Delta r\simeq120~\mathrm{m}$ (see Fig.~\ref{fig:envelope_scaling}).
The solution becomes a step function in the limit of very low values of $f_a$. For fixed central pressure, global structural changes introduced by the axion field can lead to variations of $\sim0.05M_\odot$ and $\sim0.5~ \text{km}$, at most. 

Consequently, Eq.~\eqref{eq:dTdP} describing the envelope structure also has to be modified by
\begin{equation}
\frac{\partial T}{\partial p} = \frac{\sigma T_s^4}{{\cal K} \rho (g + g_a)}~.
\label{eq:dTdPAxion}
\end{equation}
Thus, including the axion contribution reduces the temperature drop with pressure, leading to a higher surface temperature. Using the fact that $T_b$ is only a function of $T_s^4/g$, allows us to rescale previous results by only changing $g$ to $g+g_a$
\footnote{
We note that Gundmundsson's relation (\ref{eq:gudrel}) is
accurate within a few percent for typical temperatures. However, 
its accuracy declines when extrapolated to extremely thin envelopes, sometimes giving unphysical results where $T_s > T_b$. 
If this happens, we assume $T_b=T_s$ to recover the correct limit.}. 
That is, for a fixed $T_b$, the presence of an axion field results in a larger flux and larger effective temperature. This will lead to fast cooling when the photon luminosity dominates over the neutrino luminosity.
\begin{figure}
    \centering
    \includegraphics[width=\linewidth]{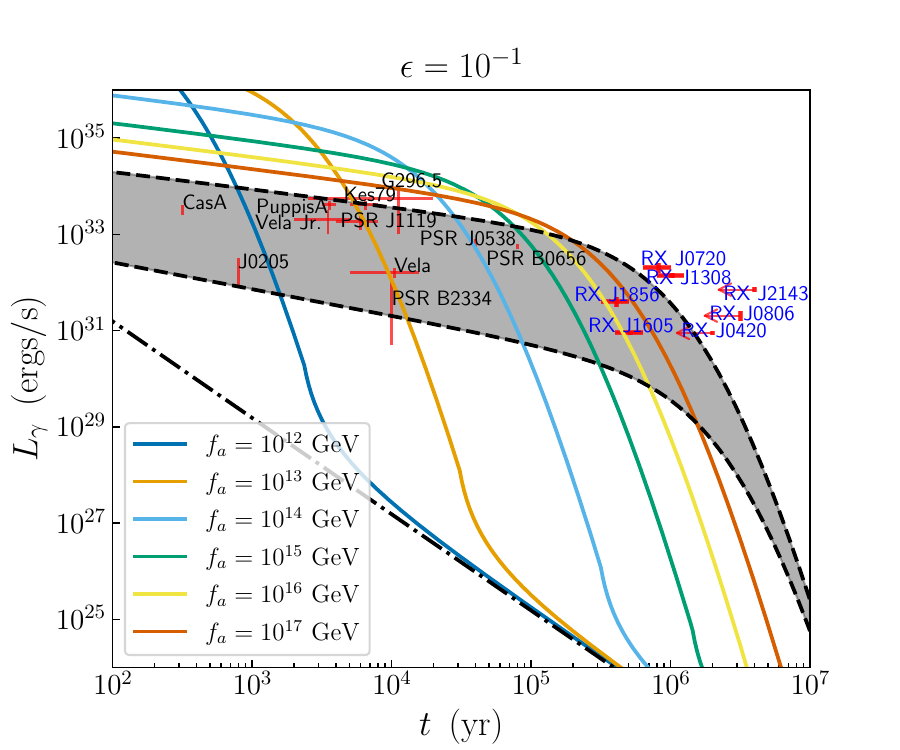}
    \includegraphics[width=\linewidth]{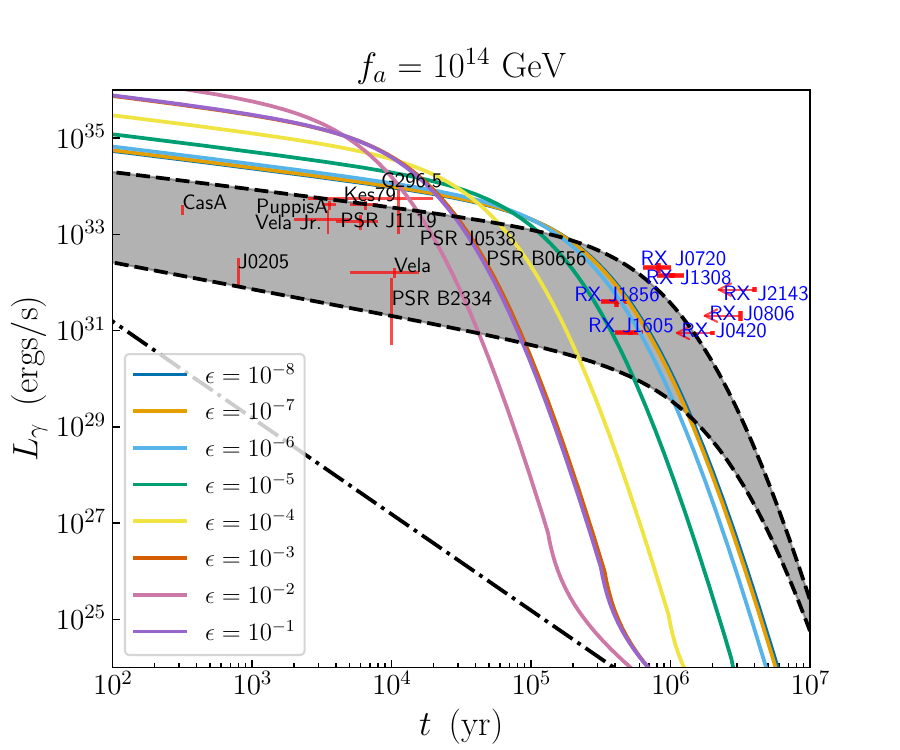}
    \caption{Comparison of NS cooling curves obtained varying $f_a$ and $\epsilon$. The gray area represents the expected variability of cooling curves in a typical NS varying the microphysics input (neutrino emissivity, superfluid gaps, or composition of the envelope) \cite{Page_2004}. The solid lines 
    show standard NS cooling curves (modified URCA only) including the effect of the presence of the axion field. The top panel shows results for $\epsilon=0.1$ and different values of $f_a$ and the bottom panel 
    results for $f_a = 10^{14}~\text{GeV}$ and different values of $\epsilon$. The observational data show a selected sample of sources (magnificent seven in blue), excluding magnetars (see \cite{Page_2004, Vigano_2013, Potekhin_2020} for a more complete sample). The dash-dotted line represents a cooling curve assuming $T_s=T_b$, which would correspond to the extremely low $f_a$ limit. }
    \label{fig:cooling_example}
\end{figure}

\textit{Neutron star cooling with axion}---The initial internal temperature of an NS born in a supernova explosion is about $T\sim10^{11}~\mathrm{K}\sim10~\mathrm{MeV}$ \cite{Pons_1999}. A few days later, the star becomes cold and transparent to neutrinos, stabilizing into an ordinary NS with an initial internal temperature of $T\sim 10^{9}~\mathrm{K}\sim0.1~\mathrm{MeV}$ \cite{Pons_2001, Pons_1999}. We focus on the cooling history of the NS from this point onward.

 After the first few years, the interior of the NS becomes isothermal [more precisely, the redshifted core temperature $T(t)\equiv T_b(r, t) e^{\nu(r)}$ becomes uniform], and we can simplify the NS cooling process as a total thermal energy balance \cite{Yakovlev_2004, Potekhin_2015} primarily driven by neutrino emission from the core and photon blackbody radiation from the surface. We account for the heat transport from the interior to the surface with the $T_s(T_b)$ relation. 
 The governing equation can be reduced to
\begin{equation}
    C_v(T)\frac{dT}{dt}=-L_{\nu}(T)-L_{\gamma}(T_s)-L_{a}+H,
    \label{cooling_eq}
\end{equation}
where $C_v(T)$ is the total heat capacity of the star and $L_{\gamma}$ and $L_{\nu}$ are the redshifted luminosities of photons and neutrinos. Typically, neutrino emission dominates for the first $10^4$ to $10^5$ yr, after which photon cooling becomes the predominant process for older NSs. 
The presence of axions can also introduce an additional cooling channel $L_a$ \cite{Sedrakian_2016, Sedrakian2019, Sedraikan_2013}, which has been employed to exclude certain parts of the axion parameter space. In this study, we do not consider this extra cooling channel, as we are focusing on the still unconstrained region of the parameter space where its impact is negligible. 
Regarding heating processes, the most significant one is arguably magnetic field decay \cite{Pons_2007b, Pons_2009, Vigano_2013}, though it is only relevant for magnetar-level field strengths. Additionally, rotochemical heating has been proposed as a potentially significant mechanism for old millisecond pulsars \cite{fernandez2005}. However, since the constraints we establish in this Letter concern the cooling of NSs with weak or moderate magnetic fields during the first million years, we will, for simplicity, neglect any heating term $H$.

The compression of the envelope caused by the presence of the axion field results in a higher $T_s$
for the same interior temperature (and therefore higher $L_{\gamma}$), causing the so-called photon 
cooling era to occur earlier. If, for typical NSs, the transition from neutrino-dominated to photon-dominated cooling occurs at core temperatures of approximately $T\approx 10^8~\text{K}$ (with a luminosity of $L_\gamma \approx 10^{32}~\text{erg/s}$), this transition happens at higher temperatures and significantly higher luminosity for NSs with higher effective gravity. Consequently, very young NSs coexisting with a sourced axion field will cool much faster. 

This effect is illustrated in Fig.~\ref{fig:cooling_example}, where we compare the standard cooling curves for a typical NS (gray area between dashed lines) with the cooling curves for NSs influenced by axions in the NGS for various values of $\epsilon$ and $f_a$. The gray area covers the typical variability observed in cooling curves when exploring the effect of different neutrino emission processes and the composition of the envelope. 
It essentially encompasses all thermally emitting NSs typically used in cooling studies, excluding magnetars, which require additional heating (not considered in this work) \cite{Vigano_2013}. To be conservative, we compare this reference cooling curve region with simplified cooling curves obtained by solving Eq.~\eqref{cooling_eq}, considering only modified URCA neutrino emission processes (see, e.g., \cite{Bottaro:2024ugp}).
Any additional contribution to neutrino emissivity would lead to faster cooling, resulting in stricter constraints.
\begin{figure}[h]
    \centering
    \includegraphics[width=\linewidth]{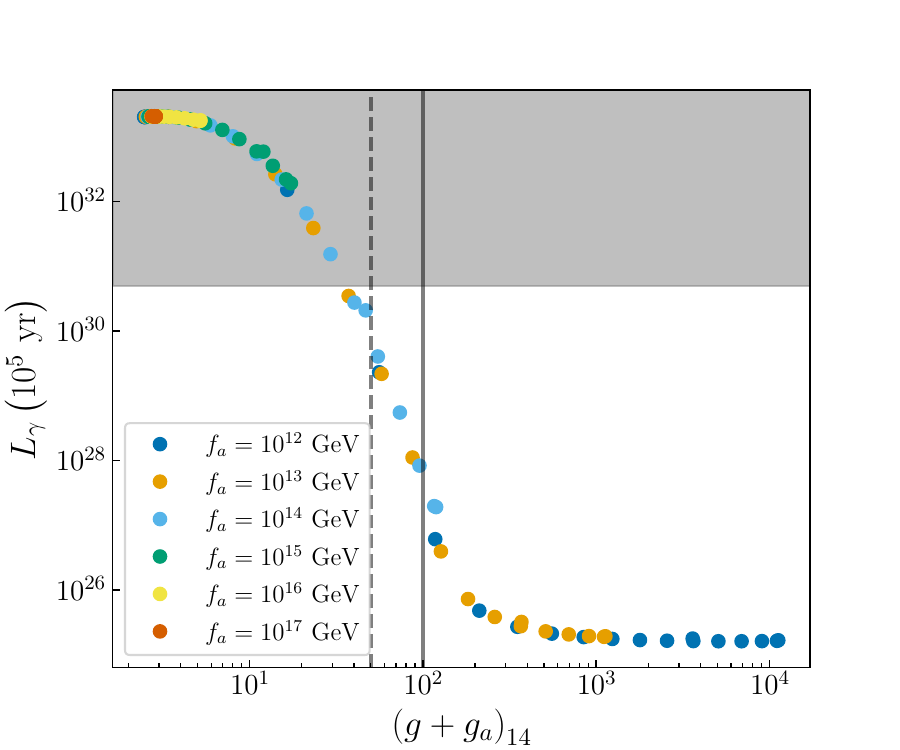}
    \caption{Photon luminosity at $t\sim 10^{5}~\mathrm{yr}$ for different $\epsilon$ and $f_a$ as a function of $\left(g + g_a\right)_{14}$. Points with the same color represent different values of $\epsilon$ for the same $f_a$. The gray shaded area corresponds to $L_{\gamma}\left(10^{5}~\mathrm{yr}\right)>5\times10^{30}~\mathrm{ergs/s}$. The vertical dashed line corresponds to $\left(g+g_a\right)_{14}=50$ and the solid one to $\left(g+g_a\right)_{14}=100$. Models with larger effective gravity reduce the envelope size too much to be consistent with the magnificent seven observations. $L_{\gamma}\left(10^{5}~\mathrm{yr}\right)$ is computed in the standard cooling approach \cite{Page_2004}. We consider these calculations an upper bound to $L_{\gamma}(10^{5}~\mathrm{yr})$ since modifications to the standard cooling scenario would make cooling faster.}
    \label{fig:L1e5}
\end{figure}
An initial examination of the figure indicates that for $\epsilon = 0.1$ and $f_a<10^{15}~\text{GeV}$, as a result of the compressed envelope caused by the presence of the axion field, we predict a much lower luminosity in middle age NSs, inconsistent with almost all observed sources, particularly the so-called ``magnificent seven": a group of NSs aged around $10^6~\text{yrs}$ and luminosities of $10^{31-33}~\text{erg/s}$ \cite{Posselt2007}. 
We emphasize that we include only modified URCA neutrino processes, which is the most conservative case. A more realistic simulation would predict even lower luminosities, especially during the first $10^3-10^4$ yr (see, e.g., \cite{Aguilera_2008} and references therein for a list
of all the relevant neutrino processes).

\begin{figure}[h]
    \centering
    \includegraphics[width=\linewidth]{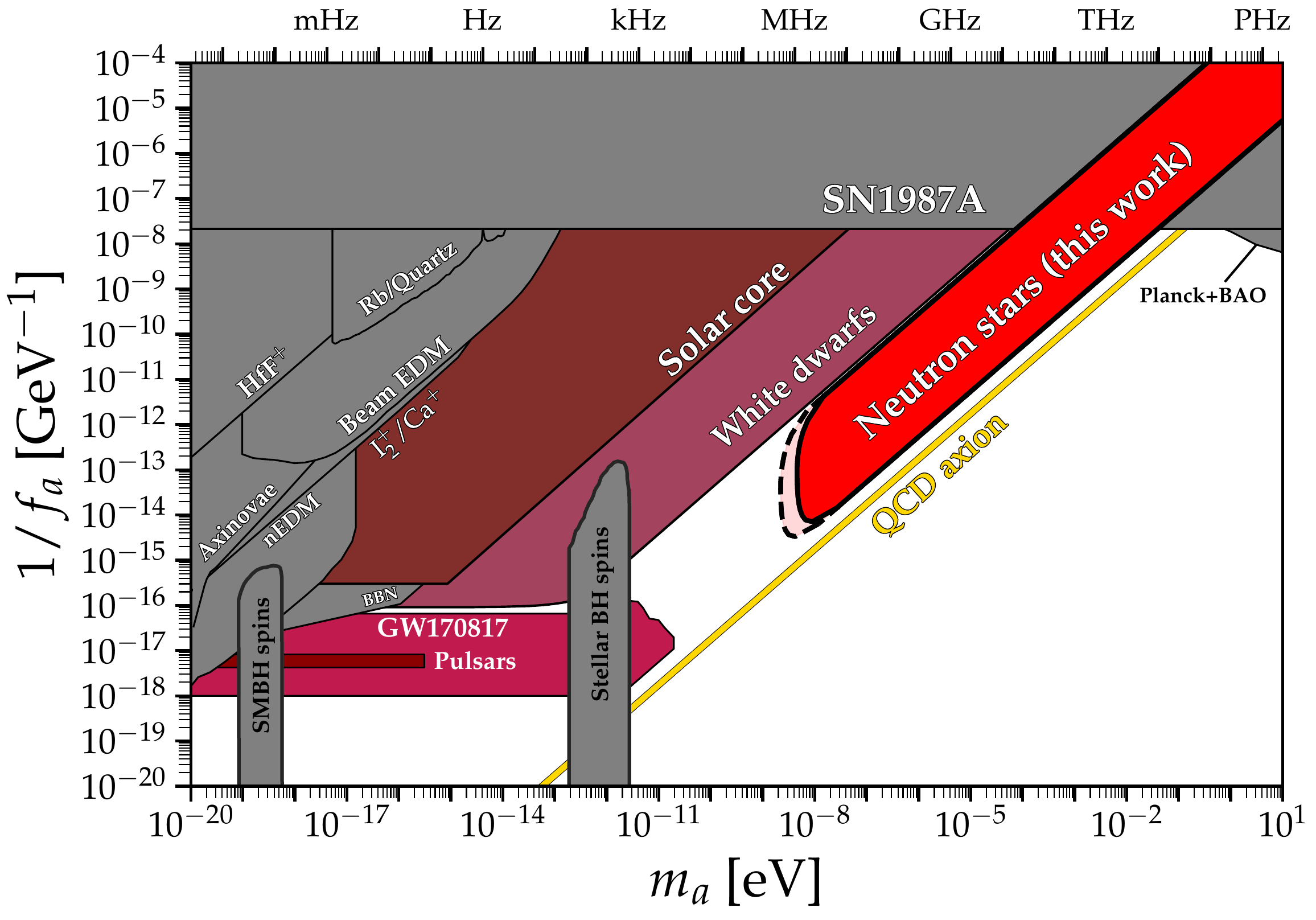}
    \caption{Constraints and projections on the axion parameter space from the anomalous fast cooling due to structural changes discussed in this Letter. The region constrained by NS cooling due to a very thin (or even inexistent) envelope is shown in bright red.
    The solid line corresponds to the conservative limit $\left(g+g_a\right)_{14}>100$, the dashed line shows where $\left(g+g_a\right)_{14}>50$. 
    Other bounds relying on the axion being sourced (shades of red) were derived in \cite{Hook:2017psm,DiLuzio:2021pxd,Zhang_2021,Balkin:2022qer}. Further bounds (gray) taken from \cite{AxionLimits,Schulthess:2022pbp,Abel:2017rtm,Roussy:2020ily,Madge:2024aot,Fuchs:2024edo,JEDI:2022hxa,Zhang:2022ewz,Fox:2023xgx,Blum:2014vsa,Mehta:2020kwu,Baryakhtar:2020gao,Unal:2020jiy,Hoof:2024quk,Caloni:2022uya,Lucente:2022vuo,Springmann:2024ret}.
    }
    \label{fig:axionExclusionPlot}
\end{figure}

To establish a more elaborated exclusion criterion, we examine the photon luminosity at $t\sim 10^{5}~\mathrm{yr}$ as a function of $\left(g + g_{a}\right)_{14}$, which turns out to be the most convenient parameter. This is depicted in Fig.~\ref{fig:L1e5}.
The gray area marks the region encompassing the luminosities of the magnificent seven and other pulsars of the same age. Each point with the same color represents a different value of $\epsilon$ for the same $f_a$.
See the Appendix for a more detailed discussion of the $\epsilon$ and $f_a$ dependence of $g_a$.
If $L_{\gamma}(10^{5}~\mathrm{yr})<5\times10^{30}~\mathrm{ergs/s}$, it is significantly below all the observational data for middle age NSs, and we will consider it ruled out. 
This typically occurs for $\left(g+g_a\right)_{14}>50$. These results can be translated into an exclusion region in the $(f_a, \epsilon)$ or equivalently the $(f_a, m_a)$ parameter space. The final result of our study is summarized in Fig.~\ref{fig:axionExclusionPlot}, where the bright red band indicates the excluded area based on the NS cooling arguments presented in this Letter.

Let us explain the boundaries of the probed region. To the right of the red area, $\epsilon>0.1$, and hence the NGS is not achieved. To the left, the NGS appears at densities below $\rho_b$, having a marginal effect on the envelope.
At the lower end, structural changes in the envelope are insignificant and the thermal evolution is unaltered.

\textit{Discussion}---Let us comment on the robustness of the presented analysis.
QCD gets strongly coupled at the densities present in NSs and it is hence not possible to reliably calculate the axion potential at these densities; see, however, \cite{Balkin:2020dsr}.
We want to stress that our effects rely on the low-density phase of the heat-blanketing envelope that is well under perturbative control.
Furthermore, we are in a $f_a$ regime where sourcing will always occur at around nuclear saturation density, even in the pessimistic scenario that the large-density (deep inner core) axion potential is minimized at $a=0$ again. 

Finally, we note that the technically natural model presented in \cite{Hook:2017psm,Hook_2018b,DiLuzio:2021pxd} avoids the NGS for \hbox{$\epsilon\gtrsim 3\times10^{-8}$}, as recently shown in \cite{Balkin:2022qer}, and hence evades the bound presented in this work.
There exist other theories that achieve a light QCD axion by tuning \cite{Elahi:2023vhu,Co:2024bme}, which might change the presented constraint, but are equally motivated as the potential assumed in this Letter.
This highlights the model dependence of the presented constraint; however, we can confidently exclude any theory that realizes a (pseudo) scalar field-induced NGS at high densities.

%
%

\textit{Acknowledgements}---A. G. B. is supported by an ACIF 2023 fellowship, cofunded by Generalitat Valenciana and the European Union through the European Social Fund.
We acknowledge funding from Conselleria d'Educació, Cultura, Universitats i Ocupació de la Generalitat Valenciana, Ministerio de Ciencia e Innovaci\'on and the European Union through the grants PID2021-127495NB-I00 (MCIN/AEI/10.13039/501100011033 and EU), ASFAE/2022/026 (MCIN and NextGenerationEU PRTR-C17.I1), and the Prometeo excellence programme grant CIPROM/2022/13. 
The work of K. B. and K. S. has been supported by the Collaborative Research Center SFB1258, the Munich Institute for Astro-, Particle and BioPhysics (MIAPbP), and by the Excellence Cluster ORIGINS, which is funded by the Deutsche Forschungsgemeinschaft (DFG, German Research Foundation) under Germany´s Excellence Strategy – EXC 2094 – 39078331.
K. S. is further supported by a research grant from Mr. and Mrs. George Zbeda.

\newpage
\textit{Appendix}---NGS formation: The formation of an NGS of nuclear matter when $a=\pi f_a$ is a necessary condition for the axion field to cause envelope thinning. In this section, we explicitly show values of $\epsilon$ for which an NGS forms for BSk26.
Taking a different EOS does not qualitatively change this conclusion.
We include the axion in the EOS as described in the main text.

An NGS, as first described in \cite{Balkin:2022qer,Balkin:2023xtr}, can be found by looking at the energy per particle as a function of density. It occurs whenever the energy per particle of the $a=\pi f_a$ phase of the EOS exhibits a minimum with an energy lower than that of the $a=0$ phase, indicating that matter is more \textit{bound} when $a=\pi f_a$. We take as a reference value $930~\mathrm{MeV}$, the energy per particle at low densities and $a=0$.

\begin{figure}[h]
    \includegraphics[width=0.45\textwidth]{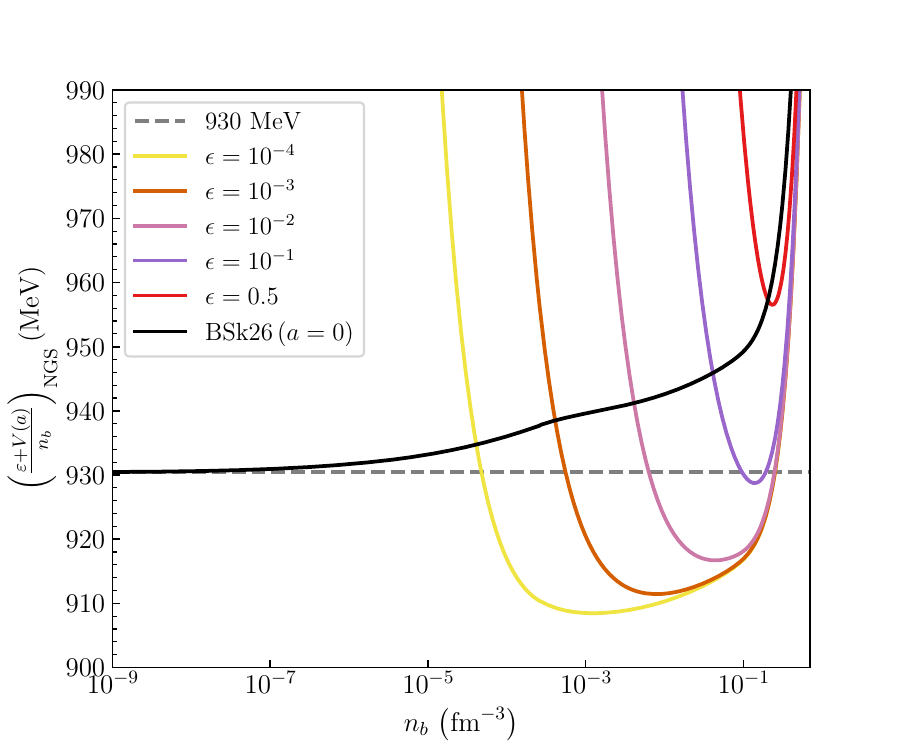}
    \includegraphics[width=0.45\textwidth]{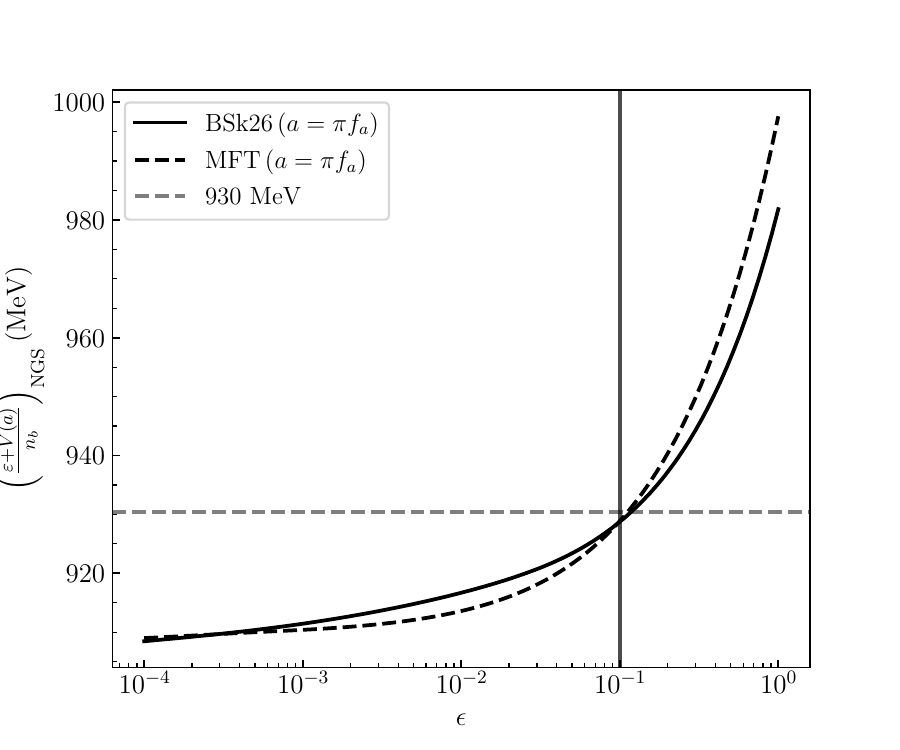}
    \caption{Upper panel: energy per particle in the two phases $a=0$ (black) and $a=\pi f_a$ (colored) for the BSk26 EOS. The dashed horizontal gray line represents the minimum energy of the ground state of nuclear matter in the $a=0$ phase, about $930~\mathrm{MeV}$ per particle.
Lower panel: minimum value of the energy per particle in the $a=\pi f_a$ phase as a function of $\epsilon$. The vertical solid line corresponds to $\epsilon=0.1$. We show that the NGS forms when $\epsilon\lesssim 0.1$ for both the relativistic mean-field model GM3 \cite{Glendenning_1985, Glendenning_1991} (black dashed line) and BSk26 (solid black line), to check that NGS formation is generic in different kinds of EOS.}
    \label{fig:NGSBSk26}
\end{figure}

In the upper panel of Fig.~\ref{fig:NGSBSk26}, we show the dependence with $n_b$ of the energy per particle for BSk26. We show results for both, the $a=0$ (black) and $a=\pi f_a$ (colored) phases, compared with the ground state energy of the $a=0$ phase. 
It is clear that the NGS forms already for $\epsilon\sim 0.1$, since the $a=\pi f_a$ phase exhibits a minimum in the energy per particle with a value lower than $930~\mathrm{MeV}$, but no longer for $\epsilon\sim 0.5$. In the lower panel, we plot the NGS energy of the $a=\pi f_a$ phase at the minimum as a function of $\epsilon$. 
The NGS will appear for all values of $\epsilon < 0.1$.

Scaling relations for $g_a$: The scaling of $g_a$ with $\left(f_a, \epsilon\right)$ across the envelope can be understood from Eq.~\eqref{eq:gphi}. 
The important factor is the axion gradient term $a'/f_a\approx 1/L$, where $L$ is the length scale over which the field decreases,
\begin{equation}
L = {\mathrm {min}}\left(\frac{1}{m_a}, R\right),
\end{equation}
where $R$ is the radius of the star.

\begin{figure}[hbt!]
    \includegraphics[width=0.45\textwidth]{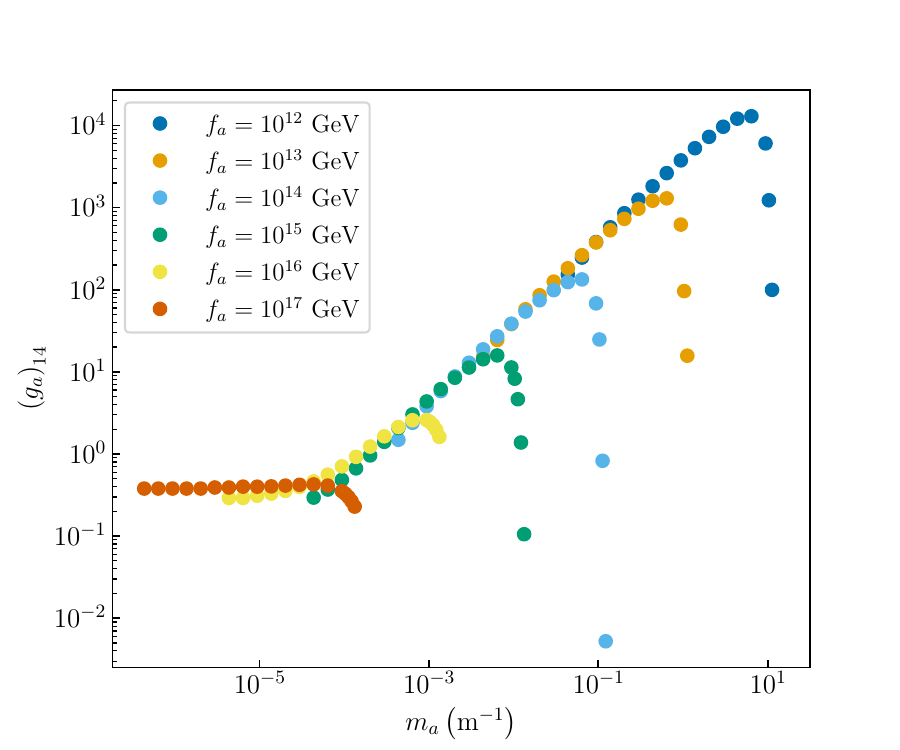}
    \caption{Axion contribution to the effective gravity as a function of $m_a$. When $m_a>10^{-4}~\mathrm{m^{-1}}$, $g_a$ scales with $m_a$.}
    \label{fig:gamean}
\end{figure}

In the region of interest, close to the star's surface, 
the axion mass is well approximated by its vacuum value, which
scales with $\sqrt{\epsilon}/f_a$,
\begin{equation}\label{eq:ma_vac}
    \left(m_a\right)_{\mathrm{vac}}=\sqrt{\epsilon}\left(\frac{m_{\pi}f_{\pi}}{f_a}\right)\frac{\sqrt{\beta}}{2}.
\end{equation}
As long as $1/m_a$ is shorter than the size of the star ($m_a \gtrsim 10^{-4}$ m$^{-1}$), we can expect the scaling
\begin{equation}\label{eq:ga_scaling}
g_a\sim \left(m_a\right)_{\mathrm{vac}} \frac{\Delta m_N^{*}}{m_N^{*}}
\propto \sqrt{\epsilon} {f_a}^{-1}.
\end{equation}
We show the effective gravity as a function of $m_a$ in Fig.~\ref{fig:gamean}.

This scaling suggests that increasing $\epsilon$ would always accelerate cooling as $g_a$ grows. 
However, this only occurs until  
$\epsilon=\epsilon^{\mathrm{NGS}}$, the value above which the NGS disappears. 
This is clearly seen in Fig.~\ref{fig:gamean}, where we observe a sudden drop of $g_a$ once the critical value of $\epsilon$ is reached. 
This also explains why the cooling at $\epsilon=10^{-2}$ is faster than at $\epsilon=10^{-1}$ in the lower panel of Fig.~\ref{fig:cooling_example}: as one approaches the end of the NGS, $g_a$ starts to drop making cooling slower again.

Figure~\ref{fig:gamean} and Eq.~\eqref{eq:ga_scaling} also allow us to understand the shape of the excluded region in Fig.~\ref{fig:axionExclusionPlot}.
To zeroth order, the exclusion can be understood as three lines: a vertical line and two diagonals.
The vertical line is fixed by the necessity of a large surface gravity $g_a\propto m_a$.
The lower and the upper diagonals are given by the need for the NGS to be present and the need for the NGS density to be larger than the envelope density.
The deviation from the vertical line is due to deviations from the simple $g_a\propto (m_a)_\text{vac}$ behavior as discussed above and finite $f_a$ effects.

\bibliography{bibliography}

\providecommand{\noopsort}[1]{}\providecommand{\singleletter}[1]{#1}%
\begin{thebibliography}{84}%
\makeatletter
\providecommand \@ifxundefined [1]{%
 \@ifx{#1\undefined}
}%
\providecommand \@ifnum [1]{%
 \ifnum #1\expandafter \@firstoftwo
 \else \expandafter \@secondoftwo
 \fi
}%
\providecommand \@ifx [1]{%
 \ifx #1\expandafter \@firstoftwo
 \else \expandafter \@secondoftwo
 \fi
}%
\providecommand \natexlab [1]{#1}%
\providecommand \enquote  [1]{``#1''}%
\providecommand \bibnamefont  [1]{#1}%
\providecommand \bibfnamefont [1]{#1}%
\providecommand \citenamefont [1]{#1}%
\providecommand \href@noop [0]{\@secondoftwo}%
\providecommand \href [0]{\begingroup \@sanitize@url \@href}%
\providecommand \@href[1]{\@@startlink{#1}\@@href}%
\providecommand \@@href[1]{\endgroup#1\@@endlink}%
\providecommand \@sanitize@url [0]{\catcode `\\12\catcode `\$12\catcode `\&12\catcode `\#12\catcode `\^12\catcode `\_12\catcode `\%12\relax}%
\providecommand \@@startlink[1]{}%
\providecommand \@@endlink[0]{}%
\providecommand \url  [0]{\begingroup\@sanitize@url \@url }%
\providecommand \@url [1]{\endgroup\@href {#1}{\urlprefix }}%
\providecommand \urlprefix  [0]{URL }%
\providecommand \Eprint [0]{\href }%
\providecommand \doibase [0]{http://dx.doi.org/}%
\providecommand \selectlanguage [0]{\@gobble}%
\providecommand \bibinfo  [0]{\@secondoftwo}%
\providecommand \bibfield  [0]{\@secondoftwo}%
\providecommand \translation [1]{[#1]}%
\providecommand \BibitemOpen [0]{}%
\providecommand \bibitemStop [0]{}%
\providecommand \bibitemNoStop [0]{.\EOS\space}%
\providecommand \EOS [0]{\spacefactor3000\relax}%
\providecommand \BibitemShut  [1]{\csname bibitem#1\endcsname}%
\let\auto@bib@innerbib\@empty
\bibitem [{\citenamefont {Peccei}\ and\ \citenamefont {Quinn}(1977{\natexlab{a}})}]{Peccei_1977a}%
  \BibitemOpen
  \bibfield  {author} {\bibinfo {author} {\bibfnamefont {R.~D.}\ \bibnamefont {Peccei}}\ and\ \bibinfo {author} {\bibfnamefont {H.~R.}\ \bibnamefont {Quinn}},\ }\href {\doibase 10.1103/PhysRevLett.38.1440} {\bibfield  {journal} {\bibinfo  {journal} {Phys. Rev. Lett.}\ }\textbf {\bibinfo {volume} {38}},\ \bibinfo {pages} {1440} (\bibinfo {year} {1977}{\natexlab{a}})}\BibitemShut {NoStop}%
\bibitem [{\citenamefont {Peccei}\ and\ \citenamefont {Quinn}(1977{\natexlab{b}})}]{Peccei_1977b}%
  \BibitemOpen
  \bibfield  {author} {\bibinfo {author} {\bibfnamefont {R.~D.}\ \bibnamefont {Peccei}}\ and\ \bibinfo {author} {\bibfnamefont {H.~R.}\ \bibnamefont {Quinn}},\ }\href {\doibase 10.1103/PhysRevD.16.1791} {\bibfield  {journal} {\bibinfo  {journal} {Phys. Rev. D}\ }\textbf {\bibinfo {volume} {16}},\ \bibinfo {pages} {1791} (\bibinfo {year} {1977}{\natexlab{b}})}\BibitemShut {NoStop}%
\bibitem [{\citenamefont {Wilczek}(1978)}]{Wilczek_1978}%
  \BibitemOpen
  \bibfield  {author} {\bibinfo {author} {\bibfnamefont {F.}~\bibnamefont {Wilczek}},\ }\href {\doibase 10.1103/PhysRevLett.40.279} {\bibfield  {journal} {\bibinfo  {journal} {Phys. Rev. Lett.}\ }\textbf {\bibinfo {volume} {40}},\ \bibinfo {pages} {279} (\bibinfo {year} {1978})}\BibitemShut {NoStop}%
\bibitem [{\citenamefont {Weinberg}(1978)}]{Weinberg_1978}%
  \BibitemOpen
  \bibfield  {author} {\bibinfo {author} {\bibfnamefont {S.}~\bibnamefont {Weinberg}},\ }\href {\doibase 10.1103/PhysRevLett.40.223} {\bibfield  {journal} {\bibinfo  {journal} {Phys. Rev. Lett.}\ }\textbf {\bibinfo {volume} {40}},\ \bibinfo {pages} {223} (\bibinfo {year} {1978})}\BibitemShut {NoStop}%
\bibitem [{\citenamefont {Raffelt}(2008)}]{Raffelt_2008}%
  \BibitemOpen
  \bibfield  {author} {\bibinfo {author} {\bibfnamefont {G.~G.}\ \bibnamefont {Raffelt}},\ }\href {\doibase 10.1007/978-3-540-73518-2_3} {\bibfield  {journal} {\bibinfo  {journal} {Lect. Notes Phys.}\ }\textbf {\bibinfo {volume} {741}},\ \bibinfo {pages} {51} (\bibinfo {year} {2008})}\BibitemShut {NoStop}%
\bibitem [{\citenamefont {Caputo}\ and\ \citenamefont {Raffelt}(2024)}]{Caputo_2024}%
  \BibitemOpen
  \bibfield  {author} {\bibinfo {author} {\bibfnamefont {A.}~\bibnamefont {Caputo}}\ and\ \bibinfo {author} {\bibfnamefont {G.}~\bibnamefont {Raffelt}},\ }\href {\doibase 10.22323/1.454.0041} {\bibfield  {journal} {\bibinfo  {journal} {PoS}\ }\textbf {\bibinfo {volume} {COSMICWISPers}},\ \bibinfo {pages} {041} (\bibinfo {year} {2024})}\BibitemShut {NoStop}%
\bibitem [{\citenamefont {Page}\ \emph {et~al.}(2011)\citenamefont {Page}, \citenamefont {Prakash}, \citenamefont {Lattimer},\ and\ \citenamefont {Steiner}}]{Page_2010}%
  \BibitemOpen
  \bibfield  {author} {\bibinfo {author} {\bibfnamefont {D.}~\bibnamefont {Page}}, \bibinfo {author} {\bibfnamefont {M.}~\bibnamefont {Prakash}}, \bibinfo {author} {\bibfnamefont {J.~M.}\ \bibnamefont {Lattimer}}, \ and\ \bibinfo {author} {\bibfnamefont {A.~W.}\ \bibnamefont {Steiner}},\ }\href {\doibase 10.1103/PhysRevLett.106.081101} {\bibfield  {journal} {\bibinfo  {journal} {Phys. Rev. Lett.}\ }\textbf {\bibinfo {volume} {106}},\ \bibinfo {pages} {081101} (\bibinfo {year} {2011})}\BibitemShut {NoStop}%
\bibitem [{\citenamefont {Fischer}\ \emph {et~al.}(2016)\citenamefont {Fischer}, \citenamefont {Chakraborty}, \citenamefont {Giannotti}, \citenamefont {Mirizzi}, \citenamefont {Payez},\ and\ \citenamefont {Ringwald}}]{Fischer_2016}%
  \BibitemOpen
  \bibfield  {author} {\bibinfo {author} {\bibfnamefont {T.}~\bibnamefont {Fischer}}, \bibinfo {author} {\bibfnamefont {S.}~\bibnamefont {Chakraborty}}, \bibinfo {author} {\bibfnamefont {M.}~\bibnamefont {Giannotti}}, \bibinfo {author} {\bibfnamefont {A.}~\bibnamefont {Mirizzi}}, \bibinfo {author} {\bibfnamefont {A.}~\bibnamefont {Payez}}, \ and\ \bibinfo {author} {\bibfnamefont {A.}~\bibnamefont {Ringwald}},\ }\href {\doibase 10.1103/PhysRevD.94.085012} {\bibfield  {journal} {\bibinfo  {journal} {Phys. Rev. D}\ }\textbf {\bibinfo {volume} {94}},\ \bibinfo {pages} {085012} (\bibinfo {year} {2016})}\BibitemShut {NoStop}%
\bibitem [{\citenamefont {Leinson}(2014)}]{Leinson_2014}%
  \BibitemOpen
  \bibfield  {author} {\bibinfo {author} {\bibfnamefont {L.~B.}\ \bibnamefont {Leinson}},\ }\href {\doibase 10.1088/1475-7516/2014/08/031} {\bibfield  {journal} {\bibinfo  {journal} {JCAP}\ }\textbf {\bibinfo {volume} {08}},\ \bibinfo {pages} {031} (\bibinfo {year} {2014})}\BibitemShut {NoStop}%
\bibitem [{\citenamefont {{Sedrakian}}(2016)}]{Sedrakian_2016}%
  \BibitemOpen
  \bibfield  {author} {\bibinfo {author} {\bibfnamefont {A.}~\bibnamefont {{Sedrakian}}},\ }\href {\doibase 10.1103/PhysRevD.93.065044} {\bibfield  {journal} {\bibinfo  {journal} {\prd}\ }\textbf {\bibinfo {volume} {93}},\ \bibinfo {eid} {065044} (\bibinfo {year} {2016})}\BibitemShut {NoStop}%
\bibitem [{\citenamefont {{Sedrakian}}(2019)}]{Sedrakian2019}%
  \BibitemOpen
  \bibfield  {author} {\bibinfo {author} {\bibfnamefont {A.}~\bibnamefont {{Sedrakian}}},\ }\href {\doibase 10.1103/PhysRevD.99.043011} {\bibfield  {journal} {\bibinfo  {journal} {\prd}\ }\textbf {\bibinfo {volume} {99}},\ \bibinfo {eid} {043011} (\bibinfo {year} {2019})}\BibitemShut {NoStop}%
\bibitem [{\citenamefont {Chang}\ \emph {et~al.}(2018)\citenamefont {Chang}, \citenamefont {Essig},\ and\ \citenamefont {McDermott}}]{Chang_2018}%
  \BibitemOpen
  \bibfield  {author} {\bibinfo {author} {\bibfnamefont {J.~H.}\ \bibnamefont {Chang}}, \bibinfo {author} {\bibfnamefont {R.}~\bibnamefont {Essig}}, \ and\ \bibinfo {author} {\bibfnamefont {S.~D.}\ \bibnamefont {McDermott}},\ }\href {\doibase 10.1007/JHEP09(2018)051} {\bibfield  {journal} {\bibinfo  {journal} {JHEP}\ }\textbf {\bibinfo {volume} {09}},\ \bibinfo {pages} {051} (\bibinfo {year} {2018})}\BibitemShut {NoStop}%
\bibitem [{\citenamefont {{Hamaguchi}}\ \emph {et~al.}(2018)\citenamefont {{Hamaguchi}}, \citenamefont {{Nagata}}, \citenamefont {{Yanagi}},\ and\ \citenamefont {{Zheng}}}]{Hamaguchi_2018}%
  \BibitemOpen
  \bibfield  {author} {\bibinfo {author} {\bibfnamefont {K.}~\bibnamefont {{Hamaguchi}}}, \bibinfo {author} {\bibfnamefont {N.}~\bibnamefont {{Nagata}}}, \bibinfo {author} {\bibfnamefont {K.}~\bibnamefont {{Yanagi}}}, \ and\ \bibinfo {author} {\bibfnamefont {J.}~\bibnamefont {{Zheng}}},\ }\href {\doibase 10.1103/PhysRevD.98.103015} {\bibfield  {journal} {\bibinfo  {journal} {\prd}\ }\textbf {\bibinfo {volume} {98}},\ \bibinfo {eid} {103015} (\bibinfo {year} {2018})}\BibitemShut {NoStop}%
\bibitem [{\citenamefont {Carenza}\ \emph {et~al.}(2019)\citenamefont {Carenza}, \citenamefont {Fischer}, \citenamefont {Giannotti}, \citenamefont {Guo}, \citenamefont {Mart\'\i{}nez-Pinedo},\ and\ \citenamefont {Mirizzi}}]{Carenza_2019}%
  \BibitemOpen
  \bibfield  {author} {\bibinfo {author} {\bibfnamefont {P.}~\bibnamefont {Carenza}}, \bibinfo {author} {\bibfnamefont {T.}~\bibnamefont {Fischer}}, \bibinfo {author} {\bibfnamefont {M.}~\bibnamefont {Giannotti}}, \bibinfo {author} {\bibfnamefont {G.}~\bibnamefont {Guo}}, \bibinfo {author} {\bibfnamefont {G.}~\bibnamefont {Mart\'\i{}nez-Pinedo}}, \ and\ \bibinfo {author} {\bibfnamefont {A.}~\bibnamefont {Mirizzi}},\ }\href {\doibase 10.1088/1475-7516/2019/10/016} {\bibfield  {journal} {\bibinfo  {journal} {JCAP}\ }\textbf {\bibinfo {volume} {10}},\ \bibinfo {pages} {016} (\bibinfo {year} {2019})},\ \bibinfo {note} {[Erratum: JCAP 05, E01 (2020)]}\BibitemShut {NoStop}%
\bibitem [{\citenamefont {Leinson}(2021)}]{Leinson_2021}%
  \BibitemOpen
  \bibfield  {author} {\bibinfo {author} {\bibfnamefont {L.~B.}\ \bibnamefont {Leinson}},\ }\href {\doibase 10.1088/1475-7516/2021/09/001} {\bibfield  {journal} {\bibinfo  {journal} {JCAP}\ }\textbf {\bibinfo {volume} {09}},\ \bibinfo {pages} {001} (\bibinfo {year} {2021})}\BibitemShut {NoStop}%
\bibitem [{\citenamefont {Buschmann}\ \emph {et~al.}(2022)\citenamefont {Buschmann}, \citenamefont {Dessert}, \citenamefont {Foster}, \citenamefont {Long},\ and\ \citenamefont {Safdi}}]{Buschmann_2021}%
  \BibitemOpen
  \bibfield  {author} {\bibinfo {author} {\bibfnamefont {M.}~\bibnamefont {Buschmann}}, \bibinfo {author} {\bibfnamefont {C.}~\bibnamefont {Dessert}}, \bibinfo {author} {\bibfnamefont {J.~W.}\ \bibnamefont {Foster}}, \bibinfo {author} {\bibfnamefont {A.~J.}\ \bibnamefont {Long}}, \ and\ \bibinfo {author} {\bibfnamefont {B.~R.}\ \bibnamefont {Safdi}},\ }\href {\doibase 10.1103/PhysRevLett.128.091102} {\bibfield  {journal} {\bibinfo  {journal} {Phys. Rev. Lett.}\ }\textbf {\bibinfo {volume} {128}},\ \bibinfo {pages} {091102} (\bibinfo {year} {2022})}\BibitemShut {NoStop}%
\bibitem [{\citenamefont {Hook}\ and\ \citenamefont {Huang}(2018)}]{Hook:2017psm}%
  \BibitemOpen
  \bibfield  {author} {\bibinfo {author} {\bibfnamefont {A.}~\bibnamefont {Hook}}\ and\ \bibinfo {author} {\bibfnamefont {J.}~\bibnamefont {Huang}},\ }\href {\doibase 10.1007/JHEP06(2018)036} {\bibfield  {journal} {\bibinfo  {journal} {JHEP}\ }\textbf {\bibinfo {volume} {06}},\ \bibinfo {pages} {036} (\bibinfo {year} {2018})}\BibitemShut {NoStop}%
\bibitem [{\citenamefont {Balkin}\ \emph {et~al.}(2020)\citenamefont {Balkin}, \citenamefont {Serra}, \citenamefont {Springmann},\ and\ \citenamefont {Weiler}}]{Balkin:2020dsr}%
  \BibitemOpen
  \bibfield  {author} {\bibinfo {author} {\bibfnamefont {R.}~\bibnamefont {Balkin}}, \bibinfo {author} {\bibfnamefont {J.}~\bibnamefont {Serra}}, \bibinfo {author} {\bibfnamefont {K.}~\bibnamefont {Springmann}}, \ and\ \bibinfo {author} {\bibfnamefont {A.}~\bibnamefont {Weiler}},\ }\href {\doibase 10.1007/JHEP07(2020)221} {\bibfield  {journal} {\bibinfo  {journal} {JHEP}\ }\textbf {\bibinfo {volume} {07}},\ \bibinfo {pages} {221} (\bibinfo {year} {2020})}\BibitemShut {NoStop}%
\bibitem [{\citenamefont {Balkin}\ \emph {et~al.}(2024)\citenamefont {Balkin}, \citenamefont {Serra}, \citenamefont {Springmann}, \citenamefont {Stelzl},\ and\ \citenamefont {Weiler}}]{Balkin:2022qer}%
  \BibitemOpen
  \bibfield  {author} {\bibinfo {author} {\bibfnamefont {R.}~\bibnamefont {Balkin}}, \bibinfo {author} {\bibfnamefont {J.}~\bibnamefont {Serra}}, \bibinfo {author} {\bibfnamefont {K.}~\bibnamefont {Springmann}}, \bibinfo {author} {\bibfnamefont {S.}~\bibnamefont {Stelzl}}, \ and\ \bibinfo {author} {\bibfnamefont {A.}~\bibnamefont {Weiler}},\ }\href {\doibase 10.1103/PhysRevD.109.095032} {\bibfield  {journal} {\bibinfo  {journal} {Phys. Rev. D}\ }\textbf {\bibinfo {volume} {109}},\ \bibinfo {pages} {095032} (\bibinfo {year} {2024})}\BibitemShut {NoStop}%
\bibitem [{\citenamefont {Balkin}\ \emph {et~al.}(2023{\natexlab{a}})\citenamefont {Balkin}, \citenamefont {Serra}, \citenamefont {Springmann}, \citenamefont {Stelzl},\ and\ \citenamefont {Weiler}}]{Balkin:2023xtr}%
  \BibitemOpen
  \bibfield  {author} {\bibinfo {author} {\bibfnamefont {R.}~\bibnamefont {Balkin}}, \bibinfo {author} {\bibfnamefont {J.}~\bibnamefont {Serra}}, \bibinfo {author} {\bibfnamefont {K.}~\bibnamefont {Springmann}}, \bibinfo {author} {\bibfnamefont {S.}~\bibnamefont {Stelzl}}, \ and\ \bibinfo {author} {\bibfnamefont {A.}~\bibnamefont {Weiler}},\ }\href@noop {} {\  (\bibinfo {year} {2023}{\natexlab{a}})},\ \Eprint {http://arxiv.org/abs/2307.14418} {arXiv:2307.14418 [hep-ph]} \BibitemShut {NoStop}%
\bibitem [{\citenamefont {{di Cortona}}\ \emph {et~al.}(2016)\citenamefont {{di Cortona}}, \citenamefont {{Hardy}}, \citenamefont {{Vega}},\ and\ \citenamefont {{Villadoro}}}]{Cortona_2016}%
  \BibitemOpen
  \bibfield  {author} {\bibinfo {author} {\bibfnamefont {G.~G.}\ \bibnamefont {{di Cortona}}}, \bibinfo {author} {\bibfnamefont {E.}~\bibnamefont {{Hardy}}}, \bibinfo {author} {\bibfnamefont {J.~P.}\ \bibnamefont {{Vega}}}, \ and\ \bibinfo {author} {\bibfnamefont {G.}~\bibnamefont {{Villadoro}}},\ }\href {\doibase 10.1007/JHEP01(2016)034} {\bibfield  {journal} {\bibinfo  {journal} {Journal of High Energy Physics}\ }\textbf {\bibinfo {volume} {2016}},\ \bibinfo {eid} {34} (\bibinfo {year} {2016})}\BibitemShut {NoStop}%
\bibitem [{\citenamefont {{Hook}}(2018)}]{Hook_2018b}%
  \BibitemOpen
  \bibfield  {author} {\bibinfo {author} {\bibfnamefont {A.}~\bibnamefont {{Hook}}},\ }\href {\doibase 10.1103/PhysRevLett.120.261802} {\bibfield  {journal} {\bibinfo  {journal} {\prl}\ }\textbf {\bibinfo {volume} {120}},\ \bibinfo {eid} {261802} (\bibinfo {year} {2018})}\BibitemShut {NoStop}%
\bibitem [{\citenamefont {Di~Luzio}\ \emph {et~al.}(2021)\citenamefont {Di~Luzio}, \citenamefont {Gavela}, \citenamefont {Quilez},\ and\ \citenamefont {Ringwald}}]{DiLuzio:2021pxd}%
  \BibitemOpen
  \bibfield  {author} {\bibinfo {author} {\bibfnamefont {L.}~\bibnamefont {Di~Luzio}}, \bibinfo {author} {\bibfnamefont {B.}~\bibnamefont {Gavela}}, \bibinfo {author} {\bibfnamefont {P.}~\bibnamefont {Quilez}}, \ and\ \bibinfo {author} {\bibfnamefont {A.}~\bibnamefont {Ringwald}},\ }\href {\doibase 10.1007/JHEP05(2021)184} {\bibfield  {journal} {\bibinfo  {journal} {JHEP}\ }\textbf {\bibinfo {volume} {05}},\ \bibinfo {pages} {184} (\bibinfo {year} {2021})}\BibitemShut {NoStop}%
\bibitem [{\citenamefont {Springmann}\ \emph {et~al.}(2024{\natexlab{a}})\citenamefont {Springmann}, \citenamefont {Stadlbauer}, \citenamefont {Stelzl},\ and\ \citenamefont {Weiler}}]{Springmann:2024mjp}%
  \BibitemOpen
  \bibfield  {author} {\bibinfo {author} {\bibfnamefont {K.}~\bibnamefont {Springmann}}, \bibinfo {author} {\bibfnamefont {M.}~\bibnamefont {Stadlbauer}}, \bibinfo {author} {\bibfnamefont {S.}~\bibnamefont {Stelzl}}, \ and\ \bibinfo {author} {\bibfnamefont {A.}~\bibnamefont {Weiler}},\ }\href@noop {} {\  (\bibinfo {year} {2024}{\natexlab{a}})},\ \Eprint {http://arxiv.org/abs/2410.10945} {arXiv:2410.10945 [hep-ph]} \BibitemShut {NoStop}%
\bibitem [{\citenamefont {Raffelt}(1996)}]{raffelt1996stars}%
  \BibitemOpen
  \bibfield  {author} {\bibinfo {author} {\bibfnamefont {G.}~\bibnamefont {Raffelt}},\ }\href {https://books.google.es/books?id=6esnbt7BfIwC} {\emph {\bibinfo {title} {Stars as Laboratories for Fundamental Physics: The Astrophysics of Neutrinos, Axions, and Other Weakly Interacting Particles}}},\ Theoretical Astrophysics\ (\bibinfo  {publisher} {University of Chicago Press},\ \bibinfo {year} {1996})\BibitemShut {NoStop}%
\bibitem [{\citenamefont {Balkin}\ \emph {et~al.}(2023{\natexlab{b}})\citenamefont {Balkin}, \citenamefont {Serra}, \citenamefont {Springmann}, \citenamefont {Stelzl},\ and\ \citenamefont {Weiler}}]{Balkin:2021zfd}%
  \BibitemOpen
  \bibfield  {author} {\bibinfo {author} {\bibfnamefont {R.}~\bibnamefont {Balkin}}, \bibinfo {author} {\bibfnamefont {J.}~\bibnamefont {Serra}}, \bibinfo {author} {\bibfnamefont {K.}~\bibnamefont {Springmann}}, \bibinfo {author} {\bibfnamefont {S.}~\bibnamefont {Stelzl}}, \ and\ \bibinfo {author} {\bibfnamefont {A.}~\bibnamefont {Weiler}},\ }\href {\doibase 10.21468/SciPostPhys.14.4.071} {\bibfield  {journal} {\bibinfo  {journal} {SciPost Phys.}\ }\textbf {\bibinfo {volume} {14}},\ \bibinfo {pages} {071} (\bibinfo {year} {2023}{\natexlab{b}})}\BibitemShut {NoStop}%
\bibitem [{\citenamefont {Balkin}\ \emph {et~al.}(2022)\citenamefont {Balkin}, \citenamefont {Serra}, \citenamefont {Springmann}, \citenamefont {Stelzl},\ and\ \citenamefont {Weiler}}]{Balkin:2021wea}%
  \BibitemOpen
  \bibfield  {author} {\bibinfo {author} {\bibfnamefont {R.}~\bibnamefont {Balkin}}, \bibinfo {author} {\bibfnamefont {J.}~\bibnamefont {Serra}}, \bibinfo {author} {\bibfnamefont {K.}~\bibnamefont {Springmann}}, \bibinfo {author} {\bibfnamefont {S.}~\bibnamefont {Stelzl}}, \ and\ \bibinfo {author} {\bibfnamefont {A.}~\bibnamefont {Weiler}},\ }\href {\doibase 10.1007/JHEP06(2022)023} {\bibfield  {journal} {\bibinfo  {journal} {JHEP}\ }\textbf {\bibinfo {volume} {06}},\ \bibinfo {pages} {023} (\bibinfo {year} {2022})}\BibitemShut {NoStop}%
\bibitem [{\citenamefont {{Anzuini}}\ \emph {et~al.}(2024)\citenamefont {{Anzuini}}, \citenamefont {{G{\'o}mez-Ba{\~n}{\'o}n}}, \citenamefont {{Pons}}, \citenamefont {{Melatos}},\ and\ \citenamefont {{Lasky}}}]{Anzuini_2024}%
  \BibitemOpen
  \bibfield  {author} {\bibinfo {author} {\bibfnamefont {F.}~\bibnamefont {{Anzuini}}}, \bibinfo {author} {\bibfnamefont {A.}~\bibnamefont {{G{\'o}mez-Ba{\~n}{\'o}n}}}, \bibinfo {author} {\bibfnamefont {J.~A.}\ \bibnamefont {{Pons}}}, \bibinfo {author} {\bibfnamefont {A.}~\bibnamefont {{Melatos}}}, \ and\ \bibinfo {author} {\bibfnamefont {P.~D.}\ \bibnamefont {{Lasky}}},\ }\href {\doibase 10.1103/PhysRevD.109.083030} {\bibfield  {journal} {\bibinfo  {journal} {\prd}\ }\textbf {\bibinfo {volume} {109}},\ \bibinfo {eid} {083030} (\bibinfo {year} {2024})}\BibitemShut {NoStop}%
\bibitem [{\citenamefont {{Gao}}\ and\ \citenamefont {{Stebbins}}(2022)}]{Gao_2022}%
  \BibitemOpen
  \bibfield  {author} {\bibinfo {author} {\bibfnamefont {C.}~\bibnamefont {{Gao}}}\ and\ \bibinfo {author} {\bibfnamefont {A.}~\bibnamefont {{Stebbins}}},\ }\href {\doibase 10.1088/1475-7516/2022/07/025} {\bibfield  {journal} {\bibinfo  {journal} {\jcap}\ }\textbf {\bibinfo {volume} {2022}},\ \bibinfo {eid} {025} (\bibinfo {year} {2022})}\BibitemShut {NoStop}%
\bibitem [{\citenamefont {Ramadan}\ \emph {et~al.}(2024)\citenamefont {Ramadan}, \citenamefont {Sakstein},\ and\ \citenamefont {Croon}}]{Ramadan:2024vfc}%
  \BibitemOpen
  \bibfield  {author} {\bibinfo {author} {\bibfnamefont {O.~F.}\ \bibnamefont {Ramadan}}, \bibinfo {author} {\bibfnamefont {J.}~\bibnamefont {Sakstein}}, \ and\ \bibinfo {author} {\bibfnamefont {D.}~\bibnamefont {Croon}},\ }\href@noop {} {\  (\bibinfo {year} {2024})},\ \Eprint {http://arxiv.org/abs/2408.02294} {arXiv:2408.02294 [hep-ph]} \BibitemShut {NoStop}%
\bibitem [{\citenamefont {{Huang}}\ \emph {et~al.}(2019)\citenamefont {{Huang}}, \citenamefont {{Johnson}}, \citenamefont {{Sagunski}}, \citenamefont {{Sakellariadou}},\ and\ \citenamefont {{Zhang}}}]{Huang_2019}%
  \BibitemOpen
  \bibfield  {author} {\bibinfo {author} {\bibfnamefont {J.}~\bibnamefont {{Huang}}}, \bibinfo {author} {\bibfnamefont {M.~C.}\ \bibnamefont {{Johnson}}}, \bibinfo {author} {\bibfnamefont {L.}~\bibnamefont {{Sagunski}}}, \bibinfo {author} {\bibfnamefont {M.}~\bibnamefont {{Sakellariadou}}}, \ and\ \bibinfo {author} {\bibfnamefont {J.}~\bibnamefont {{Zhang}}},\ }\href {\doibase 10.1103/PhysRevD.99.063013} {\bibfield  {journal} {\bibinfo  {journal} {\prd}\ }\textbf {\bibinfo {volume} {99}},\ \bibinfo {eid} {063013} (\bibinfo {year} {2019})}\BibitemShut {NoStop}%
\bibitem [{\citenamefont {{Tsuruta}}(1965)}]{Tsuruta_1965}%
  \BibitemOpen
  \bibfield  {author} {\bibinfo {author} {\bibfnamefont {S.}~\bibnamefont {{Tsuruta}}},\ }\href {\doibase 10.1038/207364a0} {\bibfield  {journal} {\bibinfo  {journal} {\nat}\ }\textbf {\bibinfo {volume} {207}},\ \bibinfo {pages} {364} (\bibinfo {year} {1965})}\BibitemShut {NoStop}%
\bibitem [{\citenamefont {{Bahcall}}\ and\ \citenamefont {{Wolf}}(1965)}]{Bahcall_1965}%
  \BibitemOpen
  \bibfield  {author} {\bibinfo {author} {\bibfnamefont {J.~N.}\ \bibnamefont {{Bahcall}}}\ and\ \bibinfo {author} {\bibfnamefont {R.~A.}\ \bibnamefont {{Wolf}}},\ }\href {\doibase 10.1103/PhysRev.140.B1452} {\bibfield  {journal} {\bibinfo  {journal} {Physical Review}\ }\textbf {\bibinfo {volume} {140}},\ \bibinfo {pages} {B1452} (\bibinfo {year} {1965})}\BibitemShut {NoStop}%
\bibitem [{\citenamefont {{Gursky}}\ \emph {et~al.}(1963)\citenamefont {{Gursky}}, \citenamefont {{Giacconi}}, \citenamefont {{Paolini}},\ and\ \citenamefont {{Rossi}}}]{Gursky_1963}%
  \BibitemOpen
  \bibfield  {author} {\bibinfo {author} {\bibfnamefont {H.}~\bibnamefont {{Gursky}}}, \bibinfo {author} {\bibfnamefont {R.}~\bibnamefont {{Giacconi}}}, \bibinfo {author} {\bibfnamefont {F.~R.}\ \bibnamefont {{Paolini}}}, \ and\ \bibinfo {author} {\bibfnamefont {B.~B.}\ \bibnamefont {{Rossi}}},\ }\href {\doibase 10.1103/PhysRevLett.11.530} {\bibfield  {journal} {\bibinfo  {journal} {\prl}\ }\textbf {\bibinfo {volume} {11}},\ \bibinfo {pages} {530} (\bibinfo {year} {1963})}\BibitemShut {NoStop}%
\bibitem [{\citenamefont {{Pons}}\ and\ \citenamefont {{Vigan{\`o}}}(2019)}]{Pons_2019}%
  \BibitemOpen
  \bibfield  {author} {\bibinfo {author} {\bibfnamefont {J.~A.}\ \bibnamefont {{Pons}}}\ and\ \bibinfo {author} {\bibfnamefont {D.}~\bibnamefont {{Vigan{\`o}}}},\ }\href {\doibase 10.1007/s41115-019-0006-7} {\bibfield  {journal} {\bibinfo  {journal} {Living Reviews in Computational Astrophysics}\ }\textbf {\bibinfo {volume} {5}},\ \bibinfo {eid} {3} (\bibinfo {year} {2019})}\BibitemShut {NoStop}%
\bibitem [{\citenamefont {{De Grandis}}\ \emph {et~al.}(2021)\citenamefont {{De Grandis}}, \citenamefont {{Taverna}}, \citenamefont {{Turolla}}, \citenamefont {{Gnarini}}, \citenamefont {{Popov}}, \citenamefont {{Zane}},\ and\ \citenamefont {{Wood}}}]{DeGrandis_2021}%
  \BibitemOpen
  \bibfield  {author} {\bibinfo {author} {\bibfnamefont {D.}~\bibnamefont {{De Grandis}}}, \bibinfo {author} {\bibfnamefont {R.}~\bibnamefont {{Taverna}}}, \bibinfo {author} {\bibfnamefont {R.}~\bibnamefont {{Turolla}}}, \bibinfo {author} {\bibfnamefont {A.}~\bibnamefont {{Gnarini}}}, \bibinfo {author} {\bibfnamefont {S.~B.}\ \bibnamefont {{Popov}}}, \bibinfo {author} {\bibfnamefont {S.}~\bibnamefont {{Zane}}}, \ and\ \bibinfo {author} {\bibfnamefont {T.~S.}\ \bibnamefont {{Wood}}},\ }\href {\doibase 10.3847/1538-4357/abfdac} {\bibfield  {journal} {\bibinfo  {journal} {\apj}\ }\textbf {\bibinfo {volume} {914}},\ \bibinfo {eid} {118} (\bibinfo {year} {2021})}\BibitemShut {NoStop}%
\bibitem [{\citenamefont {Ascenzi}\ \emph {et~al.}(2024)\citenamefont {Ascenzi}, \citenamefont {Vigan\`o}, \citenamefont {Dehman}, \citenamefont {Pons}, \citenamefont {Rea},\ and\ \citenamefont {Perna}}]{Ascenzi_2024}%
  \BibitemOpen
  \bibfield  {author} {\bibinfo {author} {\bibfnamefont {S.}~\bibnamefont {Ascenzi}}, \bibinfo {author} {\bibfnamefont {D.}~\bibnamefont {Vigan\`o}}, \bibinfo {author} {\bibfnamefont {C.}~\bibnamefont {Dehman}}, \bibinfo {author} {\bibfnamefont {J.~A.}\ \bibnamefont {Pons}}, \bibinfo {author} {\bibfnamefont {N.}~\bibnamefont {Rea}}, \ and\ \bibinfo {author} {\bibfnamefont {R.}~\bibnamefont {Perna}},\ }\href {\doibase 10.1093/mnras/stae1749} {\bibfield  {journal} {\bibinfo  {journal} {Mon. Not. Roy. Astron. Soc.}\ }\textbf {\bibinfo {volume} {533}},\ \bibinfo {pages} {201} (\bibinfo {year} {2024})}\BibitemShut {NoStop}%
\bibitem [{\citenamefont {{Yakovlev}}\ and\ \citenamefont {{Pethick}}(2004)}]{Yakovlev_2004}%
  \BibitemOpen
  \bibfield  {author} {\bibinfo {author} {\bibfnamefont {D.~G.}\ \bibnamefont {{Yakovlev}}}\ and\ \bibinfo {author} {\bibfnamefont {C.~J.}\ \bibnamefont {{Pethick}}},\ }\href {\doibase 10.1146/annurev.astro.42.053102.134013} {\bibfield  {journal} {\bibinfo  {journal} {\araa}\ }\textbf {\bibinfo {volume} {42}},\ \bibinfo {pages} {169} (\bibinfo {year} {2004})}\BibitemShut {NoStop}%
\bibitem [{\citenamefont {{Page}}\ \emph {et~al.}(2004)\citenamefont {{Page}}, \citenamefont {{Lattimer}}, \citenamefont {{Prakash}},\ and\ \citenamefont {{Steiner}}}]{Page_2004}%
  \BibitemOpen
  \bibfield  {author} {\bibinfo {author} {\bibfnamefont {D.}~\bibnamefont {{Page}}}, \bibinfo {author} {\bibfnamefont {J.~M.}\ \bibnamefont {{Lattimer}}}, \bibinfo {author} {\bibfnamefont {M.}~\bibnamefont {{Prakash}}}, \ and\ \bibinfo {author} {\bibfnamefont {A.~W.}\ \bibnamefont {{Steiner}}},\ }\href {\doibase 10.1086/424844} {\bibfield  {journal} {\bibinfo  {journal} {\apjs}\ }\textbf {\bibinfo {volume} {155}},\ \bibinfo {pages} {623} (\bibinfo {year} {2004})}\BibitemShut {NoStop}%
\bibitem [{\citenamefont {Potekhin}\ \emph {et~al.}(2015)\citenamefont {Potekhin}, \citenamefont {Pons},\ and\ \citenamefont {Page}}]{Potekhin_2015}%
  \BibitemOpen
  \bibfield  {author} {\bibinfo {author} {\bibfnamefont {A.~Y.}\ \bibnamefont {Potekhin}}, \bibinfo {author} {\bibfnamefont {J.~A.}\ \bibnamefont {Pons}}, \ and\ \bibinfo {author} {\bibfnamefont {D.}~\bibnamefont {Page}},\ }\href {\doibase 10.1007/s11214-015-0180-9} {\bibfield  {journal} {\bibinfo  {journal} {Space Sci. Rev.}\ }\textbf {\bibinfo {volume} {191}},\ \bibinfo {pages} {239} (\bibinfo {year} {2015})}\BibitemShut {NoStop}%
\bibitem [{\citenamefont {{Vigan\`o}}\ \emph {et~al.}(2013)\citenamefont {{Vigan\`o}}, \citenamefont {{Rea}}, \citenamefont {{Pons}}, \citenamefont {{Perna}}, \citenamefont {{Aguilera}},\ and\ \citenamefont {{Miralles}}}]{Vigano_2013}%
  \BibitemOpen
  \bibfield  {author} {\bibinfo {author} {\bibfnamefont {D.}~\bibnamefont {{Vigan\`o}}}, \bibinfo {author} {\bibfnamefont {N.}~\bibnamefont {{Rea}}}, \bibinfo {author} {\bibfnamefont {J.}~\bibnamefont {{Pons}}}, \bibinfo {author} {\bibfnamefont {R.}~\bibnamefont {{Perna}}}, \bibinfo {author} {\bibfnamefont {D.}~\bibnamefont {{Aguilera}}}, \ and\ \bibinfo {author} {\bibfnamefont {J.}~\bibnamefont {{Miralles}}},\ }\href {\doibase 10.1093/mnras/stt1008} {\bibfield  {journal} {\bibinfo  {journal} {\mnras}\ }\textbf {\bibinfo {volume} {434}},\ \bibinfo {pages} {123} (\bibinfo {year} {2013})}\BibitemShut {NoStop}%
\bibitem [{\citenamefont {{Kurpas}}\ \emph {et~al.}(2023)\citenamefont {{Kurpas}}, \citenamefont {{Schwope}}, \citenamefont {{Pires}}, \citenamefont {{Haberl}},\ and\ \citenamefont {{Buckley}}}]{Kurpas_2023}%
  \BibitemOpen
  \bibfield  {author} {\bibinfo {author} {\bibfnamefont {J.}~\bibnamefont {{Kurpas}}}, \bibinfo {author} {\bibfnamefont {A.~D.}\ \bibnamefont {{Schwope}}}, \bibinfo {author} {\bibfnamefont {A.~M.}\ \bibnamefont {{Pires}}}, \bibinfo {author} {\bibfnamefont {F.}~\bibnamefont {{Haberl}}}, \ and\ \bibinfo {author} {\bibfnamefont {D.~A.~H.}\ \bibnamefont {{Buckley}}},\ }\href {\doibase 10.1051/0004-6361/202346375} {\bibfield  {journal} {\bibinfo  {journal} {\aap}\ }\textbf {\bibinfo {volume} {674}},\ \bibinfo {eid} {A155} (\bibinfo {year} {2023})}\BibitemShut {NoStop}%
\bibitem [{\citenamefont {{Marino}}\ \emph {et~al.}(2024)\citenamefont {{Marino}}, \citenamefont {{Dehman}}, \citenamefont {{Kovlakas}}, \citenamefont {{Rea}}, \citenamefont {{Pons}},\ and\ \citenamefont {{Vigan{\`o}}}}]{Marino_2024}%
  \BibitemOpen
  \bibfield  {author} {\bibinfo {author} {\bibfnamefont {A.}~\bibnamefont {{Marino}}}, \bibinfo {author} {\bibfnamefont {C.}~\bibnamefont {{Dehman}}}, \bibinfo {author} {\bibfnamefont {K.}~\bibnamefont {{Kovlakas}}}, \bibinfo {author} {\bibfnamefont {N.}~\bibnamefont {{Rea}}}, \bibinfo {author} {\bibfnamefont {J.~A.}\ \bibnamefont {{Pons}}}, \ and\ \bibinfo {author} {\bibfnamefont {D.}~\bibnamefont {{Vigan{\`o}}}},\ }\href {\doibase 10.1038/s41550-024-02291-y} {\bibfield  {journal} {\bibinfo  {journal} {Nature Astronomy}\ }\textbf {\bibinfo {volume} {8}},\ \bibinfo {pages} {1020} (\bibinfo {year} {2024})}\BibitemShut {NoStop}%
\bibitem [{\citenamefont {{Beznogov}}\ \emph {et~al.}(2021)\citenamefont {{Beznogov}}, \citenamefont {{Potekhin}},\ and\ \citenamefont {{Yakovlev}}}]{Beznogov_2021}%
  \BibitemOpen
  \bibfield  {author} {\bibinfo {author} {\bibfnamefont {M.~V.}\ \bibnamefont {{Beznogov}}}, \bibinfo {author} {\bibfnamefont {A.~Y.}\ \bibnamefont {{Potekhin}}}, \ and\ \bibinfo {author} {\bibfnamefont {D.~G.}\ \bibnamefont {{Yakovlev}}},\ }\href {\doibase 10.1016/j.physrep.2021.03.004} {\bibfield  {journal} {\bibinfo  {journal} {\physrep}\ }\textbf {\bibinfo {volume} {919}},\ \bibinfo {pages} {1} (\bibinfo {year} {2021})}\BibitemShut {NoStop}%
\bibitem [{\citenamefont {Witten}(1984)}]{Witten_1984}%
  \BibitemOpen
  \bibfield  {author} {\bibinfo {author} {\bibfnamefont {E.}~\bibnamefont {Witten}},\ }\href {\doibase 10.1103/PhysRevD.30.272} {\bibfield  {journal} {\bibinfo  {journal} {Phys. Rev. D}\ }\textbf {\bibinfo {volume} {30}},\ \bibinfo {pages} {272} (\bibinfo {year} {1984})}\BibitemShut {NoStop}%
\bibitem [{\citenamefont {{Blaschke}}\ \emph {et~al.}(2001)\citenamefont {{Blaschke}}, \citenamefont {{Grigorian}},\ and\ \citenamefont {{Voskresensky}}}]{Blaschke2001}%
  \BibitemOpen
  \bibfield  {author} {\bibinfo {author} {\bibfnamefont {D.}~\bibnamefont {{Blaschke}}}, \bibinfo {author} {\bibfnamefont {H.}~\bibnamefont {{Grigorian}}}, \ and\ \bibinfo {author} {\bibfnamefont {D.~N.}\ \bibnamefont {{Voskresensky}}},\ }\href {\doibase 10.1051/0004-6361:20010005} {\bibfield  {journal} {\bibinfo  {journal} {\aap}\ }\textbf {\bibinfo {volume} {368}},\ \bibinfo {pages} {561} (\bibinfo {year} {2001})}\BibitemShut {NoStop}%
\bibitem [{\citenamefont {{Gudmundsson}}\ \emph {et~al.}(1982)\citenamefont {{Gudmundsson}}, \citenamefont {{Pethick}},\ and\ \citenamefont {{Epstein}}}]{Gudmundsson_1982}%
  \BibitemOpen
  \bibfield  {author} {\bibinfo {author} {\bibfnamefont {E.~H.}\ \bibnamefont {{Gudmundsson}}}, \bibinfo {author} {\bibfnamefont {C.~J.}\ \bibnamefont {{Pethick}}}, \ and\ \bibinfo {author} {\bibfnamefont {R.~I.}\ \bibnamefont {{Epstein}}},\ }\href {\doibase 10.1086/183840} {\bibfield  {journal} {\bibinfo  {journal} {\apjl}\ }\textbf {\bibinfo {volume} {259}},\ \bibinfo {pages} {L19} (\bibinfo {year} {1982})}\BibitemShut {NoStop}%
\bibitem [{\citenamefont {{Gudmundsson}}\ \emph {et~al.}(1983)\citenamefont {{Gudmundsson}}, \citenamefont {{Pethick}},\ and\ \citenamefont {{Epstein}}}]{Gudmundsson_1983}%
  \BibitemOpen
  \bibfield  {author} {\bibinfo {author} {\bibfnamefont {E.~H.}\ \bibnamefont {{Gudmundsson}}}, \bibinfo {author} {\bibfnamefont {C.~J.}\ \bibnamefont {{Pethick}}}, \ and\ \bibinfo {author} {\bibfnamefont {R.~I.}\ \bibnamefont {{Epstein}}},\ }\href {\doibase 10.1086/161292} {\bibfield  {journal} {\bibinfo  {journal} {\apj}\ }\textbf {\bibinfo {volume} {272}},\ \bibinfo {pages} {286} (\bibinfo {year} {1983})}\BibitemShut {NoStop}%
\bibitem [{\citenamefont {Goriely}\ \emph {et~al.}(2013)\citenamefont {Goriely}, \citenamefont {Chamel},\ and\ \citenamefont {Pearson}}]{BSk26_2013}%
  \BibitemOpen
  \bibfield  {author} {\bibinfo {author} {\bibfnamefont {S.}~\bibnamefont {Goriely}}, \bibinfo {author} {\bibfnamefont {N.}~\bibnamefont {Chamel}}, \ and\ \bibinfo {author} {\bibfnamefont {J.~M.}\ \bibnamefont {Pearson}},\ }\href {\doibase 10.1103/PhysRevC.88.061302} {\bibfield  {journal} {\bibinfo  {journal} {Phys. Rev. C}\ }\textbf {\bibinfo {volume} {88}},\ \bibinfo {pages} {061302} (\bibinfo {year} {2013})}\BibitemShut {NoStop}%
\bibitem [{\citenamefont {Tolman}(1939)}]{Tolman_1939}%
  \BibitemOpen
  \bibfield  {author} {\bibinfo {author} {\bibfnamefont {R.~C.}\ \bibnamefont {Tolman}},\ }\href {\doibase 10.1103/PhysRev.55.364} {\bibfield  {journal} {\bibinfo  {journal} {Phys. Rev.}\ }\textbf {\bibinfo {volume} {55}},\ \bibinfo {pages} {364} (\bibinfo {year} {1939})}\BibitemShut {NoStop}%
\bibitem [{\citenamefont {Oppenheimer}\ and\ \citenamefont {Volkoff}(1939)}]{Oppenheimer_1939}%
  \BibitemOpen
  \bibfield  {author} {\bibinfo {author} {\bibfnamefont {J.~R.}\ \bibnamefont {Oppenheimer}}\ and\ \bibinfo {author} {\bibfnamefont {G.~M.}\ \bibnamefont {Volkoff}},\ }\href {\doibase 10.1103/PhysRev.55.374} {\bibfield  {journal} {\bibinfo  {journal} {Phys. Rev.}\ }\textbf {\bibinfo {volume} {55}},\ \bibinfo {pages} {374} (\bibinfo {year} {1939})}\BibitemShut {NoStop}%
\bibitem [{Note1()}]{Note1}%
  \BibitemOpen
  \bibinfo {note} {We note that Gundmundsson's relation (\ref {eq:gudrel}) is accurate within a few percent for typical temperatures. However, its accuracy declines when extrapolated to extremely thin envelopes, sometimes giving unphysical results where $T_s > T_b$. If this happens, we assume $T_b=T_s$ to recover the correct limit.}\BibitemShut {Stop}%
\bibitem [{\citenamefont {Potekhin}\ \emph {et~al.}(2020)\citenamefont {Potekhin}, \citenamefont {Zyuzin}, \citenamefont {Yakovlev}, \citenamefont {Beznogov},\ and\ \citenamefont {Shibanov}}]{Potekhin_2020}%
  \BibitemOpen
  \bibfield  {author} {\bibinfo {author} {\bibfnamefont {A.~Y.}\ \bibnamefont {Potekhin}}, \bibinfo {author} {\bibfnamefont {D.~A.}\ \bibnamefont {Zyuzin}}, \bibinfo {author} {\bibfnamefont {D.~G.}\ \bibnamefont {Yakovlev}}, \bibinfo {author} {\bibfnamefont {M.~V.}\ \bibnamefont {Beznogov}}, \ and\ \bibinfo {author} {\bibfnamefont {Y.~A.}\ \bibnamefont {Shibanov}},\ }\href {\doibase 10.1093/mnras/staa1871} {\bibfield  {journal} {\bibinfo  {journal} {Monthly Notices of the Royal Astronomical Society}\ }\textbf {\bibinfo {volume} {496}},\ \bibinfo {pages} {5052–5071} (\bibinfo {year} {2020})}\BibitemShut {NoStop}%
\bibitem [{\citenamefont {{Pons}}\ \emph {et~al.}(1999)\citenamefont {{Pons}}, \citenamefont {{Reddy}}, \citenamefont {{Prakash}}, \citenamefont {{Lattimer}},\ and\ \citenamefont {{Miralles}}}]{Pons_1999}%
  \BibitemOpen
  \bibfield  {author} {\bibinfo {author} {\bibfnamefont {J.~A.}\ \bibnamefont {{Pons}}}, \bibinfo {author} {\bibfnamefont {S.}~\bibnamefont {{Reddy}}}, \bibinfo {author} {\bibfnamefont {M.}~\bibnamefont {{Prakash}}}, \bibinfo {author} {\bibfnamefont {J.~M.}\ \bibnamefont {{Lattimer}}}, \ and\ \bibinfo {author} {\bibfnamefont {J.~A.}\ \bibnamefont {{Miralles}}},\ }\href {\doibase 10.1086/306889} {\bibfield  {journal} {\bibinfo  {journal} {\apj}\ }\textbf {\bibinfo {volume} {513}},\ \bibinfo {pages} {780} (\bibinfo {year} {1999})}\BibitemShut {NoStop}%
\bibitem [{\citenamefont {{Pons}}\ \emph {et~al.}(2001)\citenamefont {{Pons}}, \citenamefont {{Miralles}}, \citenamefont {{Prakash}},\ and\ \citenamefont {{Lattimer}}}]{Pons_2001}%
  \BibitemOpen
  \bibfield  {author} {\bibinfo {author} {\bibfnamefont {J.~A.}\ \bibnamefont {{Pons}}}, \bibinfo {author} {\bibfnamefont {J.~A.}\ \bibnamefont {{Miralles}}}, \bibinfo {author} {\bibfnamefont {M.}~\bibnamefont {{Prakash}}}, \ and\ \bibinfo {author} {\bibfnamefont {J.~M.}\ \bibnamefont {{Lattimer}}},\ }\href {\doibase 10.1086/320642} {\bibfield  {journal} {\bibinfo  {journal} {\apj}\ }\textbf {\bibinfo {volume} {553}},\ \bibinfo {pages} {382} (\bibinfo {year} {2001})}\BibitemShut {NoStop}%
\bibitem [{\citenamefont {{Keller}}\ and\ \citenamefont {{Sedrakian}}(2013)}]{Sedraikan_2013}%
  \BibitemOpen
  \bibfield  {author} {\bibinfo {author} {\bibfnamefont {J.}~\bibnamefont {{Keller}}}\ and\ \bibinfo {author} {\bibfnamefont {A.}~\bibnamefont {{Sedrakian}}},\ }\href {\doibase 10.1016/j.nuclphysa.2012.11.004} {\bibfield  {journal} {\bibinfo  {journal} {\nphysa}\ }\textbf {\bibinfo {volume} {897}},\ \bibinfo {pages} {62} (\bibinfo {year} {2013})}\BibitemShut {NoStop}%
\bibitem [{\citenamefont {{Pons}}\ \emph {et~al.}(2007)\citenamefont {{Pons}}, \citenamefont {{Link}}, \citenamefont {{Miralles}},\ and\ \citenamefont {{Geppert}}}]{Pons_2007b}%
  \BibitemOpen
  \bibfield  {author} {\bibinfo {author} {\bibfnamefont {J.~A.}\ \bibnamefont {{Pons}}}, \bibinfo {author} {\bibfnamefont {B.}~\bibnamefont {{Link}}}, \bibinfo {author} {\bibfnamefont {J.~A.}\ \bibnamefont {{Miralles}}}, \ and\ \bibinfo {author} {\bibfnamefont {U.}~\bibnamefont {{Geppert}}},\ }\href {\doibase 10.1103/PhysRevLett.98.071101} {\bibfield  {journal} {\bibinfo  {journal} {\prl}\ }\textbf {\bibinfo {volume} {98}},\ \bibinfo {eid} {071101} (\bibinfo {year} {2007})}\BibitemShut {NoStop}%
\bibitem [{\citenamefont {{Pons}}\ \emph {et~al.}(2009)\citenamefont {{Pons}}, \citenamefont {{Miralles}},\ and\ \citenamefont {{Geppert}}}]{Pons_2009}%
  \BibitemOpen
  \bibfield  {author} {\bibinfo {author} {\bibfnamefont {J.~A.}\ \bibnamefont {{Pons}}}, \bibinfo {author} {\bibfnamefont {J.~A.}\ \bibnamefont {{Miralles}}}, \ and\ \bibinfo {author} {\bibfnamefont {U.}~\bibnamefont {{Geppert}}},\ }\href {\doibase 10.1051/0004-6361:200811229} {\bibfield  {journal} {\bibinfo  {journal} {\aap}\ }\textbf {\bibinfo {volume} {496}},\ \bibinfo {pages} {207} (\bibinfo {year} {2009})}\BibitemShut {NoStop}%
\bibitem [{\citenamefont {{Fern{\'a}ndez}}\ and\ \citenamefont {{Reisenegger}}(2005)}]{fernandez2005}%
  \BibitemOpen
  \bibfield  {author} {\bibinfo {author} {\bibfnamefont {R.}~\bibnamefont {{Fern{\'a}ndez}}}\ and\ \bibinfo {author} {\bibfnamefont {A.}~\bibnamefont {{Reisenegger}}},\ }\href {\doibase 10.1086/429551} {\bibfield  {journal} {\bibinfo  {journal} {\apj}\ }\textbf {\bibinfo {volume} {625}},\ \bibinfo {pages} {291} (\bibinfo {year} {2005})}\BibitemShut {NoStop}%
\bibitem [{\citenamefont {{Bottaro}}\ \emph {et~al.}(2024)\citenamefont {{Bottaro}}, \citenamefont {{Caputo}},\ and\ \citenamefont {{Fiorillo}}}]{Bottaro:2024ugp}%
  \BibitemOpen
  \bibfield  {author} {\bibinfo {author} {\bibfnamefont {S.}~\bibnamefont {{Bottaro}}}, \bibinfo {author} {\bibfnamefont {A.}~\bibnamefont {{Caputo}}}, \ and\ \bibinfo {author} {\bibfnamefont {D.~F.~G.}\ \bibnamefont {{Fiorillo}}},\ }\href {\doibase 10.1088/1475-7516/2024/11/015} {\bibfield  {journal} {\bibinfo  {journal} {\jcap}\ }\textbf {\bibinfo {volume} {2024}},\ \bibinfo {eid} {015} (\bibinfo {year} {2024})}\BibitemShut {NoStop}%
\bibitem [{\citenamefont {{Posselt}}\ \emph {et~al.}(2007)\citenamefont {{Posselt}}, \citenamefont {{Popov}}, \citenamefont {{Haberl}}, \citenamefont {{Tr{\"u}mper}}, \citenamefont {{Turolla}},\ and\ \citenamefont {{Neuh{\"a}user}}}]{Posselt2007}%
  \BibitemOpen
  \bibfield  {author} {\bibinfo {author} {\bibfnamefont {B.}~\bibnamefont {{Posselt}}}, \bibinfo {author} {\bibfnamefont {S.~B.}\ \bibnamefont {{Popov}}}, \bibinfo {author} {\bibfnamefont {F.}~\bibnamefont {{Haberl}}}, \bibinfo {author} {\bibfnamefont {J.}~\bibnamefont {{Tr{\"u}mper}}}, \bibinfo {author} {\bibfnamefont {R.}~\bibnamefont {{Turolla}}}, \ and\ \bibinfo {author} {\bibfnamefont {R.}~\bibnamefont {{Neuh{\"a}user}}},\ }\href {\doibase 10.1007/s10509-007-9344-8} {\bibfield  {journal} {\bibinfo  {journal} {\apss}\ }\textbf {\bibinfo {volume} {308}},\ \bibinfo {pages} {171} (\bibinfo {year} {2007})}\BibitemShut {NoStop}%
\bibitem [{\citenamefont {{Aguilera}}\ \emph {et~al.}(2008)\citenamefont {{Aguilera}}, \citenamefont {{Pons}},\ and\ \citenamefont {{Miralles}}}]{Aguilera_2008}%
  \BibitemOpen
  \bibfield  {author} {\bibinfo {author} {\bibfnamefont {D.~N.}\ \bibnamefont {{Aguilera}}}, \bibinfo {author} {\bibfnamefont {J.~A.}\ \bibnamefont {{Pons}}}, \ and\ \bibinfo {author} {\bibfnamefont {J.~A.}\ \bibnamefont {{Miralles}}},\ }\href {\doibase 10.1051/0004-6361:20078786} {\bibfield  {journal} {\bibinfo  {journal} {\aap}\ }\textbf {\bibinfo {volume} {486}},\ \bibinfo {pages} {255} (\bibinfo {year} {2008})}\BibitemShut {NoStop}%
\bibitem [{\citenamefont {{Zhang}}\ \emph {et~al.}(2021)\citenamefont {{Zhang}}, \citenamefont {{Lyu}}, \citenamefont {{Huang}}, \citenamefont {{Johnson}}, \citenamefont {{Sagunski}}, \citenamefont {{Sakellariadou}},\ and\ \citenamefont {{Yang}}}]{Zhang_2021}%
  \BibitemOpen
  \bibfield  {author} {\bibinfo {author} {\bibfnamefont {J.}~\bibnamefont {{Zhang}}}, \bibinfo {author} {\bibfnamefont {Z.}~\bibnamefont {{Lyu}}}, \bibinfo {author} {\bibfnamefont {J.}~\bibnamefont {{Huang}}}, \bibinfo {author} {\bibfnamefont {M.~C.}\ \bibnamefont {{Johnson}}}, \bibinfo {author} {\bibfnamefont {L.}~\bibnamefont {{Sagunski}}}, \bibinfo {author} {\bibfnamefont {M.}~\bibnamefont {{Sakellariadou}}}, \ and\ \bibinfo {author} {\bibfnamefont {H.}~\bibnamefont {{Yang}}},\ }\href {\doibase 10.1103/PhysRevLett.127.161101} {\bibfield  {journal} {\bibinfo  {journal} {\prl}\ }\textbf {\bibinfo {volume} {127}},\ \bibinfo {eid} {161101} (\bibinfo {year} {2021})}\BibitemShut {NoStop}%
\bibitem [{\citenamefont {O'Hare}(2020)}]{AxionLimits}%
  \BibitemOpen
  \bibfield  {author} {\bibinfo {author} {\bibfnamefont {C.}~\bibnamefont {O'Hare}},\ }\href {\doibase 10.5281/zenodo.3932430} {\enquote {\bibinfo {title} {cajohare/axionlimits: Axionlimits},}\ }\bibinfo {howpublished} {\url{https://cajohare.github.io/AxionLimits/}} (\bibinfo {year} {2020})\BibitemShut {NoStop}%
\bibitem [{\citenamefont {Schulthess}\ \emph {et~al.}(2022)\citenamefont {Schulthess} \emph {et~al.}}]{Schulthess:2022pbp}%
  \BibitemOpen
  \bibfield  {author} {\bibinfo {author} {\bibfnamefont {I.}~\bibnamefont {Schulthess}} \emph {et~al.},\ }\href {\doibase 10.1103/PhysRevLett.129.191801} {\bibfield  {journal} {\bibinfo  {journal} {Phys. Rev. Lett.}\ }\textbf {\bibinfo {volume} {129}},\ \bibinfo {pages} {191801} (\bibinfo {year} {2022})}\BibitemShut {NoStop}%
\bibitem [{\citenamefont {Abel}\ \emph {et~al.}(2017)\citenamefont {Abel} \emph {et~al.}}]{Abel:2017rtm}%
  \BibitemOpen
  \bibfield  {author} {\bibinfo {author} {\bibfnamefont {C.}~\bibnamefont {Abel}} \emph {et~al.},\ }\href {\doibase 10.1103/PhysRevX.7.041034} {\bibfield  {journal} {\bibinfo  {journal} {Phys. Rev. X}\ }\textbf {\bibinfo {volume} {7}},\ \bibinfo {pages} {041034} (\bibinfo {year} {2017})}\BibitemShut {NoStop}%
\bibitem [{\citenamefont {Roussy}\ \emph {et~al.}(2021)\citenamefont {Roussy} \emph {et~al.}}]{Roussy:2020ily}%
  \BibitemOpen
  \bibfield  {author} {\bibinfo {author} {\bibfnamefont {T.~S.}\ \bibnamefont {Roussy}} \emph {et~al.},\ }\href {\doibase 10.1103/PhysRevLett.126.171301} {\bibfield  {journal} {\bibinfo  {journal} {Phys. Rev. Lett.}\ }\textbf {\bibinfo {volume} {126}},\ \bibinfo {pages} {171301} (\bibinfo {year} {2021})}\BibitemShut {NoStop}%
\bibitem [{\citenamefont {Madge}\ \emph {et~al.}(2024)\citenamefont {Madge}, \citenamefont {Perez},\ and\ \citenamefont {Meir}}]{Madge:2024aot}%
  \BibitemOpen
  \bibfield  {author} {\bibinfo {author} {\bibfnamefont {E.}~\bibnamefont {Madge}}, \bibinfo {author} {\bibfnamefont {G.}~\bibnamefont {Perez}}, \ and\ \bibinfo {author} {\bibfnamefont {Z.}~\bibnamefont {Meir}},\ }\href {\doibase 10.1103/PhysRevD.110.015008} {\bibfield  {journal} {\bibinfo  {journal} {Phys. Rev. D}\ }\textbf {\bibinfo {volume} {110}},\ \bibinfo {pages} {015008} (\bibinfo {year} {2024})}\BibitemShut {NoStop}%
\bibitem [{\citenamefont {Fuchs}\ \emph {et~al.}(2024)\citenamefont {Fuchs}, \citenamefont {Kirk}, \citenamefont {Madge}, \citenamefont {Paranjape}, \citenamefont {Peik}, \citenamefont {Perez}, \citenamefont {Ratzinger},\ and\ \citenamefont {Tiedau}}]{Fuchs:2024edo}%
  \BibitemOpen
  \bibfield  {author} {\bibinfo {author} {\bibfnamefont {E.}~\bibnamefont {Fuchs}}, \bibinfo {author} {\bibfnamefont {F.}~\bibnamefont {Kirk}}, \bibinfo {author} {\bibfnamefont {E.}~\bibnamefont {Madge}}, \bibinfo {author} {\bibfnamefont {C.}~\bibnamefont {Paranjape}}, \bibinfo {author} {\bibfnamefont {E.}~\bibnamefont {Peik}}, \bibinfo {author} {\bibfnamefont {G.}~\bibnamefont {Perez}}, \bibinfo {author} {\bibfnamefont {W.}~\bibnamefont {Ratzinger}}, \ and\ \bibinfo {author} {\bibfnamefont {J.}~\bibnamefont {Tiedau}},\ }\href@noop {} {\  (\bibinfo {year} {2024})},\ \Eprint {http://arxiv.org/abs/2407.15924} {arXiv:2407.15924 [hep-ph]} \BibitemShut {NoStop}%
\bibitem [{\citenamefont {Karanth}\ \emph {et~al.}(2023)\citenamefont {Karanth} \emph {et~al.}}]{JEDI:2022hxa}%
  \BibitemOpen
  \bibfield  {author} {\bibinfo {author} {\bibfnamefont {S.}~\bibnamefont {Karanth}} \emph {et~al.} (\bibinfo {collaboration} {JEDI}),\ }\href {\doibase 10.1103/PhysRevX.13.031004} {\bibfield  {journal} {\bibinfo  {journal} {Phys. Rev. X}\ }\textbf {\bibinfo {volume} {13}},\ \bibinfo {pages} {031004} (\bibinfo {year} {2023})}\BibitemShut {NoStop}%
\bibitem [{\citenamefont {Zhang}\ \emph {et~al.}(2023)\citenamefont {Zhang}, \citenamefont {Banerjee}, \citenamefont {Leyser}, \citenamefont {Perez}, \citenamefont {Schiller}, \citenamefont {Budker},\ and\ \citenamefont {Antypas}}]{Zhang:2022ewz}%
  \BibitemOpen
  \bibfield  {author} {\bibinfo {author} {\bibfnamefont {X.}~\bibnamefont {Zhang}}, \bibinfo {author} {\bibfnamefont {A.}~\bibnamefont {Banerjee}}, \bibinfo {author} {\bibfnamefont {M.}~\bibnamefont {Leyser}}, \bibinfo {author} {\bibfnamefont {G.}~\bibnamefont {Perez}}, \bibinfo {author} {\bibfnamefont {S.}~\bibnamefont {Schiller}}, \bibinfo {author} {\bibfnamefont {D.}~\bibnamefont {Budker}}, \ and\ \bibinfo {author} {\bibfnamefont {D.}~\bibnamefont {Antypas}},\ }\href {\doibase 10.1103/PhysRevLett.130.251002} {\bibfield  {journal} {\bibinfo  {journal} {Phys. Rev. Lett.}\ }\textbf {\bibinfo {volume} {130}},\ \bibinfo {pages} {251002} (\bibinfo {year} {2023})}\BibitemShut {NoStop}%
\bibitem [{\citenamefont {Fox}\ \emph {et~al.}(2023)\citenamefont {Fox}, \citenamefont {Weiner},\ and\ \citenamefont {Xiao}}]{Fox:2023xgx}%
  \BibitemOpen
  \bibfield  {author} {\bibinfo {author} {\bibfnamefont {P.~J.}\ \bibnamefont {Fox}}, \bibinfo {author} {\bibfnamefont {N.}~\bibnamefont {Weiner}}, \ and\ \bibinfo {author} {\bibfnamefont {H.}~\bibnamefont {Xiao}},\ }\href {\doibase 10.1103/PhysRevD.108.095043} {\bibfield  {journal} {\bibinfo  {journal} {Phys. Rev. D}\ }\textbf {\bibinfo {volume} {108}},\ \bibinfo {pages} {095043} (\bibinfo {year} {2023})}\BibitemShut {NoStop}%
\bibitem [{\citenamefont {Blum}\ \emph {et~al.}(2014)\citenamefont {Blum}, \citenamefont {D'Agnolo}, \citenamefont {Lisanti},\ and\ \citenamefont {Safdi}}]{Blum:2014vsa}%
  \BibitemOpen
  \bibfield  {author} {\bibinfo {author} {\bibfnamefont {K.}~\bibnamefont {Blum}}, \bibinfo {author} {\bibfnamefont {R.~T.}\ \bibnamefont {D'Agnolo}}, \bibinfo {author} {\bibfnamefont {M.}~\bibnamefont {Lisanti}}, \ and\ \bibinfo {author} {\bibfnamefont {B.~R.}\ \bibnamefont {Safdi}},\ }\href {\doibase 10.1016/j.physletb.2014.07.059} {\bibfield  {journal} {\bibinfo  {journal} {Phys. Lett. B}\ }\textbf {\bibinfo {volume} {737}},\ \bibinfo {pages} {30} (\bibinfo {year} {2014})}\BibitemShut {NoStop}%
\bibitem [{\citenamefont {Mehta}\ \emph {et~al.}(2021)\citenamefont {Mehta}, \citenamefont {Demirtas}, \citenamefont {Long}, \citenamefont {Marsh}, \citenamefont {Mcallister},\ and\ \citenamefont {Stott}}]{Mehta:2020kwu}%
  \BibitemOpen
  \bibfield  {author} {\bibinfo {author} {\bibfnamefont {V.~M.}\ \bibnamefont {Mehta}}, \bibinfo {author} {\bibfnamefont {M.}~\bibnamefont {Demirtas}}, \bibinfo {author} {\bibfnamefont {C.}~\bibnamefont {Long}}, \bibinfo {author} {\bibfnamefont {D.~J.~E.}\ \bibnamefont {Marsh}}, \bibinfo {author} {\bibfnamefont {L.}~\bibnamefont {Mcallister}}, \ and\ \bibinfo {author} {\bibfnamefont {M.~J.}\ \bibnamefont {Stott}},\ }\href@noop {} {\  (\bibinfo {year} {2021})},\ \Eprint {http://arxiv.org/abs/2011.08693} {arXiv:2011.08693 [hep-th]} \BibitemShut {NoStop}%
\bibitem [{\citenamefont {Baryakhtar}\ \emph {et~al.}(2021)\citenamefont {Baryakhtar}, \citenamefont {Galanis}, \citenamefont {Lasenby},\ and\ \citenamefont {Simon}}]{Baryakhtar:2020gao}%
  \BibitemOpen
  \bibfield  {author} {\bibinfo {author} {\bibfnamefont {M.}~\bibnamefont {Baryakhtar}}, \bibinfo {author} {\bibfnamefont {M.}~\bibnamefont {Galanis}}, \bibinfo {author} {\bibfnamefont {R.}~\bibnamefont {Lasenby}}, \ and\ \bibinfo {author} {\bibfnamefont {O.}~\bibnamefont {Simon}},\ }\href {\doibase 10.1103/PhysRevD.103.095019} {\bibfield  {journal} {\bibinfo  {journal} {Phys. Rev. D}\ }\textbf {\bibinfo {volume} {103}},\ \bibinfo {pages} {095019} (\bibinfo {year} {2021})}\BibitemShut {NoStop}%
\bibitem [{\citenamefont {\"Unal}\ \emph {et~al.}(2021)\citenamefont {\"Unal}, \citenamefont {Pacucci},\ and\ \citenamefont {Loeb}}]{Unal:2020jiy}%
  \BibitemOpen
  \bibfield  {author} {\bibinfo {author} {\bibfnamefont {C.}~\bibnamefont {\"Unal}}, \bibinfo {author} {\bibfnamefont {F.}~\bibnamefont {Pacucci}}, \ and\ \bibinfo {author} {\bibfnamefont {A.}~\bibnamefont {Loeb}},\ }\href {\doibase 10.1088/1475-7516/2021/05/007} {\bibfield  {journal} {\bibinfo  {journal} {JCAP}\ }\textbf {\bibinfo {volume} {05}},\ \bibinfo {pages} {007} (\bibinfo {year} {2021})}\BibitemShut {NoStop}%
\bibitem [{\citenamefont {Hoof}\ \emph {et~al.}(2024)\citenamefont {Hoof}, \citenamefont {Marsh}, \citenamefont {Sisk-Reyn\'es}, \citenamefont {Matthews},\ and\ \citenamefont {Reynolds}}]{Hoof:2024quk}%
  \BibitemOpen
  \bibfield  {author} {\bibinfo {author} {\bibfnamefont {S.}~\bibnamefont {Hoof}}, \bibinfo {author} {\bibfnamefont {D.~J.~E.}\ \bibnamefont {Marsh}}, \bibinfo {author} {\bibfnamefont {J.}~\bibnamefont {Sisk-Reyn\'es}}, \bibinfo {author} {\bibfnamefont {J.~H.}\ \bibnamefont {Matthews}}, \ and\ \bibinfo {author} {\bibfnamefont {C.}~\bibnamefont {Reynolds}},\ }\href@noop {} {\  (\bibinfo {year} {2024})},\ \Eprint {http://arxiv.org/abs/2406.10337} {arXiv:2406.10337 [hep-ph]} \BibitemShut {NoStop}%
\bibitem [{\citenamefont {Caloni}\ \emph {et~al.}(2022)\citenamefont {Caloni}, \citenamefont {Gerbino}, \citenamefont {Lattanzi},\ and\ \citenamefont {Visinelli}}]{Caloni:2022uya}%
  \BibitemOpen
  \bibfield  {author} {\bibinfo {author} {\bibfnamefont {L.}~\bibnamefont {Caloni}}, \bibinfo {author} {\bibfnamefont {M.}~\bibnamefont {Gerbino}}, \bibinfo {author} {\bibfnamefont {M.}~\bibnamefont {Lattanzi}}, \ and\ \bibinfo {author} {\bibfnamefont {L.}~\bibnamefont {Visinelli}},\ }\href {\doibase 10.1088/1475-7516/2022/09/021} {\bibfield  {journal} {\bibinfo  {journal} {JCAP}\ }\textbf {\bibinfo {volume} {09}},\ \bibinfo {pages} {021} (\bibinfo {year} {2022})}\BibitemShut {NoStop}%
\bibitem [{\citenamefont {Lucente}\ \emph {et~al.}(2022)\citenamefont {Lucente}, \citenamefont {Mastrototaro}, \citenamefont {Carenza}, \citenamefont {Di~Luzio}, \citenamefont {Giannotti},\ and\ \citenamefont {Mirizzi}}]{Lucente:2022vuo}%
  \BibitemOpen
  \bibfield  {author} {\bibinfo {author} {\bibfnamefont {G.}~\bibnamefont {Lucente}}, \bibinfo {author} {\bibfnamefont {L.}~\bibnamefont {Mastrototaro}}, \bibinfo {author} {\bibfnamefont {P.}~\bibnamefont {Carenza}}, \bibinfo {author} {\bibfnamefont {L.}~\bibnamefont {Di~Luzio}}, \bibinfo {author} {\bibfnamefont {M.}~\bibnamefont {Giannotti}}, \ and\ \bibinfo {author} {\bibfnamefont {A.}~\bibnamefont {Mirizzi}},\ }\href {\doibase 10.1103/PhysRevD.105.123020} {\bibfield  {journal} {\bibinfo  {journal} {Phys. Rev. D}\ }\textbf {\bibinfo {volume} {105}},\ \bibinfo {pages} {123020} (\bibinfo {year} {2022})}\BibitemShut {NoStop}%
\bibitem [{\citenamefont {Springmann}\ \emph {et~al.}(2024{\natexlab{b}})\citenamefont {Springmann}, \citenamefont {Stadlbauer}, \citenamefont {Stelzl},\ and\ \citenamefont {Weiler}}]{Springmann:2024ret}%
  \BibitemOpen
  \bibfield  {author} {\bibinfo {author} {\bibfnamefont {K.}~\bibnamefont {Springmann}}, \bibinfo {author} {\bibfnamefont {M.}~\bibnamefont {Stadlbauer}}, \bibinfo {author} {\bibfnamefont {S.}~\bibnamefont {Stelzl}}, \ and\ \bibinfo {author} {\bibfnamefont {A.}~\bibnamefont {Weiler}},\ }\href@noop {} {\  (\bibinfo {year} {2024}{\natexlab{b}})},\ \Eprint {http://arxiv.org/abs/2410.19902} {arXiv:2410.19902 [hep-ph]} \BibitemShut {NoStop}%
\bibitem [{\citenamefont {Elahi}\ \emph {et~al.}(2023)\citenamefont {Elahi}, \citenamefont {Elor}, \citenamefont {Kivel}, \citenamefont {Laux}, \citenamefont {Najjari},\ and\ \citenamefont {Yu}}]{Elahi:2023vhu}%
  \BibitemOpen
  \bibfield  {author} {\bibinfo {author} {\bibfnamefont {F.}~\bibnamefont {Elahi}}, \bibinfo {author} {\bibfnamefont {G.}~\bibnamefont {Elor}}, \bibinfo {author} {\bibfnamefont {A.}~\bibnamefont {Kivel}}, \bibinfo {author} {\bibfnamefont {J.}~\bibnamefont {Laux}}, \bibinfo {author} {\bibfnamefont {S.}~\bibnamefont {Najjari}}, \ and\ \bibinfo {author} {\bibfnamefont {F.}~\bibnamefont {Yu}},\ }\href {\doibase 10.1103/PhysRevD.108.L031701} {\bibfield  {journal} {\bibinfo  {journal} {Phys. Rev. D}\ }\textbf {\bibinfo {volume} {108}},\ \bibinfo {pages} {L031701} (\bibinfo {year} {2023})}\BibitemShut {NoStop}%
\bibitem [{\citenamefont {{Co}}\ \emph {et~al.}(2024)\citenamefont {{Co}}, \citenamefont {{Gherghetta}}, \citenamefont {{Liu}},\ and\ \citenamefont {{Lyu}}}]{Co:2024bme}%
  \BibitemOpen
  \bibfield  {author} {\bibinfo {author} {\bibfnamefont {R.~T.}\ \bibnamefont {{Co}}}, \bibinfo {author} {\bibfnamefont {T.}~\bibnamefont {{Gherghetta}}}, \bibinfo {author} {\bibfnamefont {Z.}~\bibnamefont {{Liu}}}, \ and\ \bibinfo {author} {\bibfnamefont {K.-F.}\ \bibnamefont {{Lyu}}},\ }\href {\doibase 10.1007/JHEP09(2024)145} {\bibfield  {journal} {\bibinfo  {journal} {Journal of High Energy Physics}\ }\textbf {\bibinfo {volume} {2024}},\ \bibinfo {eid} {145} (\bibinfo {year} {2024})}\BibitemShut {NoStop}%
\bibitem [{\citenamefont {{Glendenning}}(1985)}]{Glendenning_1985}%
  \BibitemOpen
  \bibfield  {author} {\bibinfo {author} {\bibfnamefont {N.~K.}\ \bibnamefont {{Glendenning}}},\ }\href {\doibase 10.1086/163253} {\bibfield  {journal} {\bibinfo  {journal} {\apj}\ }\textbf {\bibinfo {volume} {293}},\ \bibinfo {pages} {470} (\bibinfo {year} {1985})}\BibitemShut {NoStop}%
\bibitem [{\citenamefont {Glendenning}\ and\ \citenamefont {Moszkowski}(1991)}]{Glendenning_1991}%
  \BibitemOpen
  \bibfield  {author} {\bibinfo {author} {\bibfnamefont {N.~K.}\ \bibnamefont {Glendenning}}\ and\ \bibinfo {author} {\bibfnamefont {S.~A.}\ \bibnamefont {Moszkowski}},\ }\href {\doibase 10.1103/PhysRevLett.67.2414} {\bibfield  {journal} {\bibinfo  {journal} {Phys. Rev. Lett.}\ }\textbf {\bibinfo {volume} {67}},\ \bibinfo {pages} {2414} (\bibinfo {year} {1991})}\BibitemShut {NoStop}%
\end{thebibliography}%

\end{document}